\documentclass[a4paper,11pt]{article}
\usepackage{jheppub_modified_green} 

\usepackage{tensor}
\usepackage{mathtools}
\usepackage{subcaption}
\usepackage{booktabs}

\usepackage{tikz}
\usetikzlibrary{hobby,calc,shapes, positioning, backgrounds,trees, decorations, decorations.markings, decorations.pathmorphing, arrows, shapes.geometric, snakes}
\tikzset{
	graviton/.style={line width=.8pt, -latex,decorate, decoration={snake, segment length=4pt,amplitude=1.8pt, pre length=.1cm, post length=.25cm}},
	worldline/.style={gray, line width=1pt},
	worldlineBold/.style={black, line width=.6pt},
	zUndirected/.style={line width=1pt},
	zParticle/.style={line width=1pt,postaction={decorate},decoration={markings,mark=at position .6 with {\arrow[#1]{latex}}}},
	zParticleF/.style={line width=1pt,postaction={decorate}},
	cscalar/.style={line width=1pt,postaction={decorate},decoration={markings,mark=at position .6 with {\arrow[#1]{latex}}}},
	cscalar2/.style={line width=1pt,postaction={decorate},decoration={markings,mark=at position .8 with {\arrow[#1]{latex}}}},
	photon/.style={line width=.8pt, decorate, decoration={snake, segment length=4pt, amplitude=1.8pt,  pre length=.1cm, post length=.1cm}},
    photon2/.style={line width=2pt, gray, decorate, decoration={snake, segment length=5pt, amplitude=1.8pt,  pre length=.1cm, post length=.1cm,}},
    cross/.style={path picture={ 
      \draw[gray,thick]
    (path picture bounding box.south east) -- (path picture bounding box.north west) (path picture bounding box.south west) -- (path picture bounding box.north east);
    }}
}

\usepackage{tikz-feynman}
\tikzfeynmanset{compat=1.1.0}
\tikzfeynmanset{
graviton/.style={circle, draw=green!60, fill=green!5, very thick, minimum size=7mm}
}
\tikzfeynmanset{
codot/.style={/tikz/shape=circle,
/tikz/fill=red,/tikz/minimum size=0.1cm,/tikz/inner sep=1.8pt}
}
\tikzfeynmanset{sb/.style=
{/tikz/shape=circle,
/tikz/fill=black,
 /tikz/minimum size=0.3cm,/tikz/inner sep=1.pt
 } }
\tikzfeynmanset{myblob/.style=
{/tikz/shape=ellipse,
/tikz/fill=red,
 /tikz/minimum width=0.5cm,
 } }
 \tikzfeynmanset{myblobFF/.style=
{/tikz/shape=circle,
/tikz/fill=black,
 /tikz/minimum width=0.25cm,
 } }
 \tikzfeynmanset{myblob2/.style=
{/tikz/shape=rectangle,
/tikz/fill=red,
 /tikz/minimum width=0.1cm,/tikz/inner sep=1.0pt
 } }
  \tikzfeynmanset{HV/.style=
{/tikz/shape=circle,
/tikz/fill={rgb:black,1;white,2},
 /tikz/minimum size=0.1cm,
 } }
  \tikzfeynmanset{GR/.style=
{/tikz/shape=ellipse,
/tikz/fill={rgb:black,1;white,2},
 /tikz/minimum size=0.3cm,
 } }
  \tikzfeynmanset{fp/.style=
{/tikz/shape=rectangle,
/tikz/fill=red,
 /tikz/minimum width=0.1cm,/tikz/inner sep=1.8pt
 } }
  \tikzfeynmanset{myvertex/.style=
{/tikz/shape=circle,
/tikz/fill=black,
 /tikz/minimum width=0.05cm,
 } }
  \tikzfeynmanset{myvertex1/.style=
{/tikz/shape=circle,
/tikz/fill=gray,
 /tikz/minimum width=0.05mm,
 } }
\tikzset{box/.pic={\filldraw[fill=black]  (0,0) circle (2.5pt);
				   \filldraw [fill=black] (0.5,0) circle (2.5pt);
			       \draw [line width=5pt] (0,0) -- (0.5,0);}}

\tikzset{wiggle/.style={decorate, decoration=snake}}

 \tikzfeynmanset{HV/.style=
{/tikz/shape=circle,
/tikz/fill={rgb:black,1;white,2},
 /tikz/minimum size=0.01cm,/tikz/inner sep=1.8pt
 } }

\DeclarePairedDelimiter\abs{\lvert}{\rvert}%
\newcommand{\metdens}{\sqrt{\abs{g}}}
\newcommand{\bmetdens}{\sqrt{\abs{\bar g}}}
\newcommand{\ee}{\mathrm{e}}
\newcommand{\ii}{\mathrm{i}}
\newcommand{\kinpar}{y}

\newcommand{\mdot}{{\cdot}}

\def\nn{\nonumber}
\usepackage{hyperref}
\hypersetup{
	colorlinks=true,
	linktoc=page,
	citecolor=americanrose,
	linkcolor=cadmiumgreen,
	urlcolor=blue} 
\definecolor{americanrose}{rgb}{1.0, 0.01, 0.24}
\definecolor{cadmiumgreen}{rgb}{0.0, 0.42, 0.24}

\title{The gravitational Compton amplitude from flat and curved spacetimes at second post-Minkowskian order  \!\!\!\!\!\!\!\!\!\!\!\!\!\!\!\!\!\!\!\!\!\!\!\!\!\!\!\!\!\!\!\!}

\author{N. Emil J. Bjerrum-Bohr$^{1,2}$,}
\author{Gang Chen$^{1,2}$,}
\author{Carl Jordan Eriksen$^{2}$,}
\author{Nabha Shah$^{1,2}$}
\affiliation{$^1$Center of Gravity, Niels Bohr Institute,
Blegdamsvej 17, DK-2100 Copenhagen, Denmark\\[-14pt]}
\affiliation{$^2$Niels Bohr International Academy, Niels Bohr Institute, University of Copenhagen,\\
Blegdamsvej 17, DK-2100 Copenhagen, Denmark\\[-10pt]}

\emailAdd{bjbohr@nbi.dk}
\emailAdd{gang.chen@nbi.ku.dk}
\emailAdd{carl.eriksen@nbi.ku.dk}
\emailAdd{nabha.shah@nbi.ku.dk}

\abstract{We utilize various computational techniques in flat and curved backgrounds to calculate the classical gravitational Compton amplitude up to the second post-Minkowskian order. Our novel result supports the use of worldline effective field theory in non-trivial background spacetimes to obtain new theoretical insights that can both enhance computational efficiency and provide useful cross-checks of results, particularly in the context of classical binary black hole mergers.}


\begin{document}

\maketitle

\tableofcontents

\flushbottom

\newpage

\section{Introduction}
Since the LIGO-Virgo collaboration first confirmed the existence of gravitational waves \cite{LIGOScientific:2016aoc}, research into the dynamics of gravitational binary systems has advanced significantly, with new developments in computational technology that utilize novel quantum field-theoretic methods for gravitational physics \cite{Iwasaki:1971iy,Goldberger:2004jt,Neill:2013wsa,Bjerrum-Bohr:2013bxa,Bjerrum-Bohr:2018xdl,Cheung:2018wkq,Kosower:2018adc}. Examples of recent studies include, for instance, the improvement of analytical descriptions of the dynamics of two-body mergers during the inspiral phase \cite{ Kalin:2020mvi,Mogull:2020sak, Cristofoli:2019neg,Bern:2019nnu,Antonelli:2019ytb,Bern:2019crd,Parra-Martinez:2020dzs,DiVecchia:2020ymx,Damour:2020tta,Kalin:2020fhe,Bern:2020buy,DiVecchia:2021ndb,DiVecchia:2021bdo,Herrmann:2021tct,Bjerrum-Bohr:2021vuf,Bjerrum-Bohr:2021din,Damgaard:2021ipf,Brandhuber:2021eyq,Bjerrum-Bohr:2021wwt,Bern:2021yeh,Dlapa:2021npj,Dlapa:2021vgp,Jakobsen:2021zvh,Jakobsen:2021lvp,Bern:2021dqo,Bjerrum-Bohr:2022blt,Bjerrum-Bohr:2022ows,Bern:2022jvn,Dlapa:2022lmu,Jakobsen:2022fcj,Adamo:2022ooq,Bern:2022kto,Luna:2023uwd,Damgaard:2023ttc,DiVecchia:2023frv,Heissenberg:2023uvo,Jakobsen:2023ndj,Bjerrum-Bohr:2024hul,Driesse:2024xad,Chen:2024mmm,Bohnenblust:2024hkw,Driesse:2024feo,Chen:2024bpf,Akpinar:2025bkt,Bjerrum-Bohr:2025lpw,Foffa:2019hrb,Blumlein:2019zku,Foffa:2019yfl,Blumlein:2020pog,Blumlein:2020znm, Kalin:2019inp, Kalin:2019rwq}  and research focused on gravitational bremsstrahlung effects \cite{Brandhuber:2023hhy,Herderschee:2023fxh,Elkhidir:2023dco,Georgoudis:2023lgf,Caron-Huot:2023vxl,Jakobsen:2023hig,Bohnenblust:2023qmy,Bini:2023fiz,Georgoudis:2023eke,Brandhuber:2023hhl,DeAngelis:2023lvf,Adamo:2024oxy,Bini:2024rsy,Alessio:2024wmz,Georgoudis:2024pdz,Brunello:2024ibk,Brandhuber:2024qdn,Bohnenblust:2025gir}.

Compton amplitudes, which describe the scattering of gravitons off a massive compact object, are essential elements in such applications \cite{
Chen:2022kpm, Aoude:2020onz,
Chung:2020rrz,Chiodaroli:2021eug,Bautista:2021wfy,Aoude:2022trd,Bjerrum-Bohr:2023jau,Chen:2022clh,Bjerrum-Bohr:2023iey,Chen:2024mlx,Cangemi:2023bpe,Bjerrum-Bohr:2024fbt,Bautista:2024emt,Correia:2024jgr,Vazquez-Holm:2025ztz,Ivanov:2022qqt,Saketh:2024juq,Caron-Huot:2025tlq}, and play a vital role in deriving observables from compact loop-level integrands based on unitarity constraints. Additionally, they provide a pathway for determining the classical dynamics of gravitational perturbations around stable, curved backgrounds, such as the spacetimes of Schwarzschild or Kerr black holes. In the S-matrix approach to gravitational binary systems, computations involve suitably defined on-shell external states in flat space. By moving from flat to curved spacetimes \cite{duff}, we can open up an avenue for further insights through the development of perturbative S-matrix methods around non-trivial classical backgrounds \cite{ Kosmopoulos:2023bwc,Cheung:2023lnj,Cheung:2024byb,PhysRevLett.132.251603, Mougiakakos:2024lif,Akpinar:2025huz,Hoogeveen:2025tew}. In support of this line of inquiry, we investigate the graviton two-point function within a framework that describes a massive gravitating body by a single worldline in a non-trivial background, following established effective field theory approaches \cite{Goldberger:2004jt, Porto:2005ac, Kalin:2020mvi, Mogull:2020sak}. To our knowledge, this specific object has only been addressed in literature to the first post-Minkowskian order, in a weak-field expansion in \cite{Comberiati:2022ldk} and in the context of curved space expansions in \cite{Kosmopoulos:2023bwc}, and partially to the second post-Minkowskian order in \cite{Akpinar:2025huz}. Additionally, the second post-Lorentzian (i.e. order $e^4$) correction to the Compton amplitude in scalar quantum electrodynamics was computed in \cite{Brunello:2024ibk}. The systematic incorporation of the recoil effects produced by deflections of the massive background-generating source, and their impact on a lighter body traversing in the resultant gravitational fields, has been discussed in \cite{Kosmopoulos:2023bwc, Cheung:2023lnj,Cheung:2024byb}. By narrowing our focus to the gravitational interactions of an isolated massive body, we independently study the effects and simplifications arising from perturbations of the Schwarzschild-Tangherlini metric, decoupled from those associated with the self-force expansion. We use dimensional regularization throughout our calculations, making the choice of a $d$-dimensional metric mandatory. Considering the computation of the Compton amplitude in this background, while initially more challenging, could lead to enhanced efficiency by leveraging resummations of effects stemming from the use of non-trivial gravitational backgrounds.

We will investigate this field-theoretic approach by conducting a systematic perturbative computation of the Compton amplitude to second post-Minkowskian order while working in both a flat and a non-trivial curved background spacetime. While we do not a priori expect to find any physical discrepancies between computational schemes when working with point-like sources, we will observe that working in a curved background while incorporating deflections of the massive body \cite{Kosmopoulos:2023bwc}, or their consequences through ``recoil operators'' 
\cite{Cheung:2023lnj}, appears to provide various benefits.   

The outline of our presentation is as follows. In section \ref{sec:Weak-field expansion} of this paper, we review the derivation of Feynman rules for worldline effective field theories in flat space. Following \cite{Kosmopoulos:2023bwc, Cheung:2023lnj,Cheung:2024byb}, \emph{curved space} Feynman rules valid for a general, asymptotically flat spacetime are established in section \ref{sec:curved-expansion}. Using these rules, section \ref{sec:comp_amp} discusses and compares the diagrammatic contributions to the Compton amplitude via curved and flat space expansions, while, in section \ref{sec:1pm} and section \ref{sec:2pm}, we explicitly compute the Compton amplitude to first and second order in a post-Minkowskian expansion, respectively. The expressions derived from flat and curved space worldline quantum field theory rules are shown to agree exactly, clarifying current discussions on the use of curved space techniques and inviting higher-order consistency checks. The result at this order displays the anticipated infrared divergent behavior consistent with the predictions of the established Weinberg soft theorem~\cite{Weinberg:1965nx}. Furthermore, in the geometric-optics limit, we recover the exact, universal expression for the bending angle from the amplitude, matching the results for a massless scalar or photon scattering off a massive particle \cite{Bjerrum-Bohr:2014zsa,Bjerrum-Bohr:2016hpa,Bai:2016ivl,Chi:2019owc}. We also provide a cross-check of the second-order result using the systematic and gauge-invariant diagrammatic framework of heavy-mass effective field theory \cite{Damgaard:2019lfh, Brandhuber:2021kpo, Brandhuber:2021eyq, Brandhuber:2021bsf}. Finally, we conclude in section \ref{sec:conc}, outlining potential applications and future research avenues. We adopt the mostly-minus metric signature everywhere and operate in natural units, where $\hbar = c = 1$.

\section{Worldline quantum field theory in flat and curved space}
\label{sec:WQFT}
We consider the gravitational interactions of a single massive body described by the Einstein-Hilbert action and a minimally coupled massive worldline action,
\begin{equation}
    \label{action}
    S[g,x] = S_\text{EH}[g] + S_\text{gf}[h] + S_\text{wl}[g,x],
\end{equation}
with
\begin{align}
    \label{actioncomp}
    S_\text{EH}[g]=-\frac{2}{\kappa^2}\int\mathrm{d}^dx\,\metdens R,  && S_\text{wl}[g,x] = -\frac{M}{2}\int\mathrm{d}\tau\,\big(g_{\mu\nu}\dot x^\mu\dot x^\nu + 1\big),
\end{align}
where $M$ is the mass of our non-spinning body and $S_\text{gf}[h]$ introduces a gauge-fixing term. Our Riemann tensor and related quantities are defined by
\begin{equation}
    R_{\mu\nu\rho\sigma} \equiv \Gamma_{\sigma\mu\rho,\nu} - \Gamma_{\sigma\nu\rho,\mu} + \Gamma\indices{^\lambda_{\nu\rho}}\Gamma_{\lambda\sigma\mu} - \Gamma\indices{^\lambda_{\mu\rho}}\Gamma_{\lambda\sigma\nu}, \quad R_{\mu\nu} \equiv R\indices{^\lambda_{\mu\lambda
u}},\quad R \equiv R\indices{^\mu_\mu},
\end{equation}
where the commas indicate partial differentiation. We note that this formalism applies to $d$-dimensional space, therefore the metric density is denoted as $\sqrt{\abs{g}}$, since $g \equiv \det{g} < 0$ for odd $d$ according to our metric signature convention. The $d$-dimensional gravitational coupling constant is taken to be
\begin{equation}
    \kappa^2 \equiv 32\pi G\tilde\mu^{2\epsilon}, \qquad \text{with} \qquad \tilde\mu^2 \equiv \frac{e^{\gamma_\text{E}}\mu^2}{4\pi},
\end{equation}
where we have opted to keep the four-dimensional definition of Newton's constant, $G$, by introducing the arbitrary mass scale, $\mu$. We will consider this action with expansions of the metric tensor around two different backgrounds. The first choice for a background metric produces the familiar \emph{weak-field expansion},
\begin{equation}
    g_{\mu\nu} = \eta_{\mu\nu} + \kappa h_{\mu\nu},
\label{eq:weak-field-exp}\end{equation}
which gives rise to standard flat space graviton interaction vertices. A natural choice of gauge in this expansion is the flat harmonic or de Donder gauge, implemented by
\begin{equation}
    S^\text{wf}_\text{gf}[h] = \eta^{\mu\nu}G^\text{wf}_\mu G^\text{wf}_\nu, \qquad \text{with} \qquad G^\text{wf}_\mu = \partial_\nu h\indices{^\nu_\mu} - \frac12\eta^{\rho\sigma}\partial_\mu h_{\rho\sigma}.
\end{equation}
For the second expansion, a general, curved background geometry is considered so that
\begin{equation}
    g_{\mu\nu} = \bar g_{\mu\nu} + \kappa h_{\mu\nu}.
\label{eq:curved-exp}\end{equation}
We shall eventually choose the background spacetime to be of the Schwarzschild-Tangherlini form \cite{Tangherlini:1963bw}. However, at this stage, we keep it generic. When working in a curved background, we choose the background-covariant harmonic gauge,
\begin{equation}
    S^\text{curv}_\text{gf}[h] = \bar g^{\mu\nu}G^\text{curv}_\mu G^\text{curv}_\nu, \qquad \text{with} \qquad G^\text{curv}_\mu = \bar g^{\nu\lambda}\bar\nabla_\nu h_{\lambda\mu} - \frac12\bar g^{\rho\sigma}\bar\nabla_\mu h_{\rho\sigma},
\end{equation}
where $\bar\nabla$ is the covariant derivative compatible with $\bar g_{\mu\nu}$.

\subsection{Weak-field expansion}
\label{sec:Weak-field expansion}
Let us briefly review standard Feynman rules for Einstein gravity in the weak-field expansion \cite{sannan1986gravity, DeWitt:1967yk,DeWitt:1967ub,badger_scattering_2024,Donoghue:2017pgk}. Inserting eq.\,\eqref {eq:weak-field-exp} into the gauge-fixed Einstein-Hilbert action and expanding to quadratic order in $h$ yields
\begin{equation}
    S_\text{EH}[\eta + \kappa h]\big\vert_{h^2} + S^\text{wf}_\text{gf}[h] = -\frac12\int\mathrm{d}^dx\, h_{\mu\nu}(x)(P^{-1})^{\mu\nu\,\rho\sigma}\partial^2h_{\rho\sigma}(x),
\end{equation}
where the inverse de Donder projector is
\begin{equation}
    (P^{-1})^{\mu\nu\,\rho\sigma} = I^{\mu\nu\,\rho\sigma} - \frac12\eta^{\mu\nu}\eta^{\rho\sigma}, \qquad I^{\mu\nu\,\rho\sigma} \equiv \eta^{\mu(\rho}\eta^{\sigma)\nu},
\label{eq:inv-de-donder}\end{equation}
and $I^{\mu\nu\,\rho\sigma}$ is the identity tensor in the space of pairs of symmetric indices. (We symmetrize and antisymmetrize with unit weight, e.g., $T^{(\mu\nu)} = \frac{1}{2}(T^{\mu\nu} + T^{\nu\mu})$.) Inverting in momentum space yields the familiar de Donder graviton propagator,
\begin{equation}
    \begin{tikzpicture}[baseline={(current bounding box.center)}]
        \coordinate (x) at (-.5,0);
        \coordinate (y) at (1.5,0);
        \draw [photon] (x) -- (y) node [midway, below] {$k$};
        \draw [fill] (x) node [above] {$h_{\mu\nu}(k)$};
        \draw [fill] (y) node [above] {$h_{\rho\sigma}(k)$};
        \end{tikzpicture} = \frac{\ii P_{\mu\nu\,\rho\sigma}}{k^2 + \ii0}, \qquad P_{\mu\nu\,\rho\sigma} \equiv I_{\mu\nu\,\rho\sigma} - \frac{1}{d-2}\eta_{\mu\nu}\eta_{\rho\sigma}.
\label{eq:de-donder}\end{equation}
For our computation, we also need the three-point and four-point graviton vertices. These vertices take their usual, lengthy form, so we do not report them here. We note only that in deriving and manipulating these expressions, along with other tensor expressions in this work, we have made extensive use of the \texttt{Mathematica} packages \texttt{xTensor} and \texttt{xPert} which are part of \texttt{xAct} \cite{xAct, Brizuela:2008ra}. We take outgoing momenta to be positive and $k_{1\cdots i} = k_1 + \cdots + k_i$.

Let us now turn our attention to the worldline action in eq.\,\eqref{actioncomp}, dropping constant terms along the way since they have no role to play. Here, we only outline the derivation of Feynman rules from this action since they have been discussed comprehensively in \cite{Mogull:2020sak}. Devoid of any interactions, the point particle traverses a classical inertial trajectory at a constant velocity, $v^\rho$. When perturbed, however, it will experience a deflection, $z^\rho(\tau)$, away from this free classical value, resulting in a background field expansion of the worldline coordinate of the form,
\begin{equation}
    x^\rho(\tau) = v^\rho\tau + z^\rho(\tau),
\label{eq:inertial-exp}\end{equation}
which, upon insertion into the worldline action, defines vertex functions involving $n$ gravitons and $m$ worldline deflections to be
\begin{equation}
   \prod_{\substack{i=1..n\\j=1..m}}\frac{\delta}{\delta h_{\mu_i\nu_i}(-k_i)}\frac{\delta}{\delta z^{\rho_j}(-\omega_j)}\ii S[h,z]\Big\vert_{h^nz^m},
\label{eq:vertex-def}\end{equation}
where the graviton, $h_{\mu\nu}$, and the deflection, $z^\rho$, should be expressed in terms of their Fourier transforms,
\begin{equation}
    h_{\mu\nu}(x) = \int_k\ee^{-\ii k\cdot x}h_{\mu\nu}(k), \qquad z^\rho(\tau) = \int_\omega\ee^{-\ii \omega\tau}z^\rho(\omega),
\end{equation}
before acting with the variations on the action. We employ the notation
\begin{equation}
    \int_k \equiv \int\frac{\mathrm{d}^dk}{(2\pi)^d}, \qquad \int_\omega \equiv \int\frac{\mathrm{d}\omega}{2\pi}.
\end{equation}
and, for convenience, also define
\begin{equation}
    \hat\delta(k) \equiv (2\pi)^d\delta^{(d)}(k), \qquad \hat\delta(\omega) \equiv 2\pi\delta(\omega).
\end{equation}
Inserting the weak-field expansion of the metric, the linearity of the worldline action implies
\begin{equation}
    S_\text{wl}[\eta + \kappa h,x] = S_\text{wl}[\eta,x] + \kappa S_\text{wl}[h,x],
\label{eq:wl-action-decomp}\end{equation}
and, expanding the worldline trajectory in the first term,
\begin{equation}
    \label{def_prop_action}
    S_\text{wl}[\eta, v\tau + z] = -\frac{M}{2}\int\mathrm{d}\tau\,\eta_{\rho\sigma}\dot z^\rho\dot z^\sigma - M\int\mathrm{d}\tau\, v\cdot\dot z.
\end{equation}
The second term in eq.\,\eqref{def_prop_action} can be dropped since it is a total derivative and we obtain the propagator for the deflection, depicted using a solid line,
\begin{equation}
    \begin{tikzpicture}[baseline={(current bounding box.center)}]
        \coordinate (in) at (-1,0);
        \coordinate (out) at (2,0);
        \coordinate (x) at (-.3,0);
        \coordinate (y) at (1.3,0);
        \draw [zUndirected] (x) -- (y) node [midway, below] {$\omega$};
        \draw [dotted, thick] (in) -- (x);
        \draw [dotted, thick] (y) -- (out);
        \draw [fill] (x) node [above] {$z^\rho(\omega)$};
        \draw [fill] (y) node [above] {$z^\sigma(\omega)$};
    \end{tikzpicture} = \frac{-\ii\eta^{\rho\sigma}}{M\omega^2 + \ii0}.
\end{equation}
The second term of eq.\,\eqref{eq:wl-action-decomp} gives rise to an infinite tower of interaction vertices on the worldline, all of which have a single graviton leg but any number of deflection legs, including none. The source of this tower can be identified by considering the Fourier transform,
\begin{equation}
    h_{\mu\nu}(x(\tau)) = \int_k\ee^{\ii k\cdot x(\tau)}h_{\mu\nu}(-k),
\end{equation}
where the expansion of the exponential after inserting the inertial expansion produces
\begin{align}
    h_{\mu\nu}(x(\tau)) &= \int_k\ee^{\ii \tau k\cdot v}\Bigg[\sum_{j=0}^\infty\frac{\ii^j(k\cdot z(\tau))^j}{j!}\Bigg]h_{\mu\nu}(-k) \notag\\
    &= \int_k\ee^{\ii \tau k\cdot v}\Bigg[\sum_{j=0}^\infty\frac{\ii^j}{j!}\prod_{l=1}^j\int_{\omega_l}\ee^{\ii \omega_l\tau}(k\cdot z(-\omega_l))\Bigg]h_{\mu\nu}(-k),
\end{align}
giving rise to, after performing the integration over $\tau$,
\begin{equation}
\begin{aligned}
    \kappa S_\text{wl}[h,x] &= -\kappa M\sum_{j=0}^\infty\frac{\ii^j}{j!}
    \int_{k,\omega_1,\ldots,\omega_j}\hat\delta(k\cdot v + \omega_{1\cdots j})
    h_{\mu\nu}(-k)\prod_{l=1}^jz^{\rho_l}(-\omega_l) \\
    &\times\Bigg[\frac12\prod_{n=1}^jk_{\rho_n}v^\mu v^\nu+
    \sum_{n=1}^j\prod_{m\neq n}^j\omega_nk_{\rho_m}\!
    v^{(\mu}\delta^{\nu)}_{\rho_n}+
    \sum_{n<m}^j\prod_{s\neq n,m}^j\omega_n\omega_mk_{\rho_s}
    \delta^{(\mu}_{\rho_n}\delta^{\nu)}_{\rho_m}\Bigg]\,,
\end{aligned}
\label{eq:wl-fouriered}\end{equation}
with
\begin{equation}
    \omega_{1\cdots j} = \sum_{i=1}^j\omega_i.
\end{equation}
As it turns out, for this paper, we require only the pieces that are zeroth and first order in the deflection. The zeroth-order piece,
\begin{equation}
    \kappa S_\text{wl}[h,x]\big\vert_{z^0} = -\frac{\kappa M}{2}\int_k\hat\delta(k\cdot v)v^\mu v^\nu h_{\mu\nu}(-k),
\label{eq:wl-source}\end{equation}
is related to the energy-momentum tensor of the Schwarzschild-Tangherlini metric, which will be discussed in the next section. It gives rise to a Feynman rule where the worldline, represented by a faint dotted line, sources a graviton,
\begin{equation}
    \begin{tikzpicture}[baseline={(current bounding box.center)}]
        \coordinate (in) at (-1,0);
        \coordinate (out) at (1,0);
        \coordinate (x) at (0,0);
        \node (k) at (0,-1.3) {$h_{\mu\nu}(k)$};
        \draw [dotted, thick] (in) -- (x);
        \draw [dotted, thick] (x) -- (out);
        \draw [photon] (x) -- (k);
        \draw [fill] (x) circle (.08);
    \end{tikzpicture} = -\frac{\ii\kappa M}{2}\hat\delta(k\cdot v)v^\mu v^\nu.
\end{equation}
We emphasize that the worldline serves only as a visual guide, that is, the above vertex has only one leg. The terms of the action which are linear in the deflection can be written as
\begin{equation}
    \kappa S_\text{wl}[h,x]\big\vert_{z^1} = -\frac{\ii\kappa M}{2}\int_k\hat\delta(k\cdot v + \omega)(v^\mu v^\nu k_\rho + 2\omega v\indices{^{(\mu}}\delta\indices*{^{\nu)}_\rho})z^\rho(-\omega)h_{\mu\nu}(-k),
\end{equation}
giving the Feynman rule,
\begin{equation}
    \begin{tikzpicture}[baseline={(current bounding box.center)}]
        \coordinate (in) at (-1,0);
        \coordinate (out) at (1,0);
        \coordinate (x) at (0,0);
        \node (k) at (0,-1.3) {$h_{\mu\nu}(k)$};
        \draw (out) node [right] {$z^\rho(\omega)$};
        \draw [dotted, thick] (in) -- (x);
        \draw [zUndirected] (x) -- (out);
        \draw [photon] (x) -- (k);
        \draw [fill] (x) circle (.08);
    \end{tikzpicture} = \frac{\kappa M}{2}\hat\delta(k\cdot v + \omega)(v^\mu v^\nu k_\rho + 2\omega v\indices{^{(\mu}}\delta\indices*{^{\nu)}_\rho}).
\end{equation}
\sloppy We will use these rules in the calculation of the Compton amplitude at first post-Minkowskian order in section \ref{sec:1pm} and second post-Minkowskian order in section \ref{sec:2pm}. First, we consider how to derive Feynman rules in a curved space expansion.

Note that in quantum field theories, defined in the bulk of spacetime, all Feynman rules are accompanied by momentum-conserving delta functions coming from the ubiquitous presence of $\int\mathrm{d}^dx$ in the action. However, since the deflection is defined only on the worldline, its Feynman rules contain merely energy-conserving delta functions. Therefore, in contrast to tree-level computations in the bulk, loop-like integrals will appear in the tree-level diagrammatic expansion of our theory due to the leftover momenta which are not constrained by momentum conservation.

\subsection{Curved space expansion}
\label{sec:curved-expansion}
We will now consider an expansion where the background field is a non-trivial position-dependent metric. By inserting this expansion into the Einstein-Hilbert action, gauge-fixed using the background-covariant harmonic gauge, and disregarding total derivatives, we find the terms linear and quadratic in $h$,
\begin{align}
    S_\text{EH}[\bar g + \kappa h]\big\vert_{h^1} &= \frac{2}{\kappa}\int\mathrm{d}^dx\,\bmetdens\bar G_{\mu\nu}h^{\mu\nu}, \label{eq:lin-eh-term}\\[1em]
    S_\text{EH}[\bar g + \kappa h]\big\vert_{h^2} + S^\text{curv}_\text{gf}[h] &= \int\mathrm{d}^dx\,\bmetdens \Bigg[\frac14\bar R(2h^{\mu\nu}h_{\mu\nu} - h^2) \notag\\&\hspace{5em}+ \bar R_{\mu\nu}(h^{\mu\nu} h - h^{\mu\lambda}h\indices{_\lambda^\nu}) - \bar R_{\rho\nu\sigma\mu}h^{\mu\nu}h^{\rho\sigma} \label{eq:quad-eh-term}\\&\hspace{5em}+ \frac12\bar\nabla_\rho h_{\mu\nu}\bar\nabla^\rho h^{\mu\nu} - \frac14\bar\nabla_\mu h\bar\nabla^\mu h\Bigg]. \notag
\end{align}
Here, we use $h \equiv \bar g^{\mu\nu}h_{\mu\nu}$ and raise and lower indices with the background metric, $\bar g_{\mu\nu}$. When we discuss the worldline action below, we will demonstrate how the contribution from eq.\,\eqref{eq:lin-eh-term} cancels when we choose $\bar g_{\mu\nu}$ to be the Schwarzschild-Tangherlini solution. For now, we concentrate on the quadratic term, splitting the action into a flat and an interacting part,
\begin{equation}
    S_\text{G}^\text{free}[h] = \eqref{eq:quad-eh-term}\big\vert_{\bar g\to\eta}, \qquad S_\text{G}^\text{int}[h]=\eqref{eq:quad-eh-term} - S_\text{G}^\text{free}[h].
\end{equation}
$S_\text{G}^\text{free}[h]$ is identical to the quadratic piece of the Einstein-Hilbert action in the weak-field expansion and gives rise to the de Donder propagator of eq.\,\eqref{eq:de-donder}. Expanding the covariant derivatives in the interaction action leaves us with
\begin{align}
    S_\text{G}^\text{int}[h] = \int\mathrm{d}^dx\,\Big[\bar{\Phi}_{[\partial^0]}^{\mu_1\nu_1\,\mu_2\nu_2\,\gamma\delta}h_{\mu_1\nu_1,\gamma}h_{\mu_2\nu_2,\delta} + \bar{\Phi}_{[\partial^1]}^{\mu_1\nu_1\,\mu_2\nu_2\,\delta}h_{\mu_1\nu_1}h_{\mu_2\nu_2,\delta} + \bar{\Phi}_{[\partial^2]}^{\mu_1\nu_1\,\mu_2\nu_2}h_{\mu_1\nu_1}h_{\mu_2\nu_2}\Big],
\label{eq:quad-act-int-short}\end{align}
where the subscript on the tensors, $\bar{\Phi}$, reflects the number of derivatives of the metric contained in them. Using\footnote{Notice that $T\indices*{^{\mu_1\nu_1\,\mu_2\nu_2}_{\rho_1\sigma_1\,\rho_2\sigma_2}}$ is defined entirely in terms of Kronecker deltas and is thus not dependent on the metric.}
\begin{equation}
    T\indices*{^{\mu_1\nu_1\,\mu_2\nu_2}_{\rho_1\sigma_1\,\rho_2\sigma_2}} \equiv I\indices{^{\mu_1(\mu_2}_{\rho_1\sigma_1}}I\indices{^{\nu_2)\nu_1}_{\rho_2\sigma_2}} - \frac12I\indices{^{\mu_1\nu_1}_{\rho_1\sigma_1}}I\indices{^{\mu_2\nu_2}_{\rho_2\sigma_2}},
\end{equation}
we can spell out
\begin{subequations}
\begin{align}
    \bar{\Phi}_{[\partial^0]}^{\mu_1\nu_1\,\mu_2\nu_2\,\gamma\delta} &= \frac18(\bmetdens\bar g^{\gamma\delta}\bar g^{\rho_1\sigma_1}\bar g^{\rho_2\sigma_2} - \eta^{\gamma\delta}\eta^{\rho_1\sigma_1}\eta^{\rho_2\sigma_2})T\indices*{^{\mu_1\nu_1\,\mu_2\nu_2}_{\rho_1\sigma_1\,\rho_2\sigma_2}}, \\
    \bar{\Phi}_{[\partial^1]}^{\mu_1\nu_1\,\mu_2\nu_2\,\delta} &= - \frac12\bmetdens\bar g^{\rho_2\sigma_2}T\indices*{^{\mu_2\nu_2\,\delta(\nu_1}_{\rho_1\sigma_1\,\rho_2\sigma_2}}\bar\Gamma^{\mu_1)\rho_1\sigma_1}, \\
    \bar{\Phi}_{[\partial^2]}^{\mu_1\nu_1\,\mu_2\nu_2} &= \operatorname{sym} P_2\frac18\bmetdens\Big(4\bar g^{\rho_1\sigma_1}\bar g^{\rho_2\sigma_2}\bar\Gamma\indices{^{\mu_1}_{\gamma\lambda}}\bar\Gamma\indices{^{\mu_2}_{\delta}^{\lambda}}T\indices*{^{\nu_1\gamma\,\nu_2\delta}_{\rho_1\sigma_1\,\rho_2\sigma_2}} + \bar R\bar g^{\rho_1\sigma_1}\bar g^{\rho_2\sigma_2}T\indices*{^{\mu_1\nu_1\,\mu_2\nu_2}_{\rho_1\sigma_1\,\rho_2\sigma_2}} \notag\\
    &\hspace{4.5em}+ 16\bar g^{\mu_1[\nu_1}\bar R^{\mu_2]\nu_2} - 8\bar R^{\mu_1\mu_2\nu_1\nu_2}\Big),
\end{align}
\end{subequations}
where $P_2$ denotes symmetrization in the index pairs, $(\mu_1,\nu_1)$ and $(\mu_2,\nu_2)$, and sym denotes symmetrization between $\mu_i$ and $\nu_i$, and with eq.\,\eqref{eq:quad-act-int-short} in hand, we can finally Fourier transform the gravitons and make the crossing symmetry of the vertex explicit, yielding
\begin{equation}
\begin{aligned}
    S_\text{G}^\text{int}[h] &= \int_{k_1,k_2}h_{\mu_1\nu_1}(-k_1)h_{\mu_2\nu_2}(-k_2)\int\mathrm{d}^dx\,\ee^{\ii(k_1+k_2)\cdot x} \\
    &\times\bigg(-\bar{\Phi}_{[\partial^0]}^{\mu_1\nu_1\,\mu_2\nu_2\,\gamma\delta}k_{1\gamma}k_{2\delta} + \frac{\ii}{2}\Big[\bar{\Phi}_{[\partial^1]}^{\mu_1\nu_1\,\mu_2\nu_2\,\delta}k_{1\delta} + \bar{\Phi}_{[\partial^1]}^{\mu_2\nu_2\,\mu_1\nu_1\,\delta}k_{2\delta}\Big] + \bar{\Phi}_{[\partial^2]}^{\mu_1\nu_1\,\mu_2\nu_2}\bigg).
\end{aligned}
\end{equation}
Defining the injected momentum, $q \equiv k_1 + k_2$, we find a vertex,
\begin{equation}
    \begin{tikzpicture}[baseline={(current bounding box.center)}]
        \coordinate (in) at (-1,0);
        \coordinate (out) at (1,0);
        \coordinate (x) at (0,0);
        \coordinate (gin) at (-1,-1.2);
        \coordinate (gout) at (1,-1.2);
        \coordinate (v) at (0,-1.2);
        \draw [dotted, thick] (in) -- (x);
        \draw [dotted, thick] (x) -- (out);
        \draw [photon2] (x) -- (v) node [black,midway,left] {$q\!\downarrow$};
        \draw [photon] (v) -- (gin) node [left, below=.5em] {$h_{\mu_1\nu_1}(k_1)$};
        \draw [photon] (v) -- (gout) node [right, below=.5em] {$h_{\mu_2\nu_2}(k_2)$};
        \node [draw,gray,thick,fill=white,circle,cross,minimum width=.3cm] at (x) {};
        \draw [fill] (v) circle (.08);
    \end{tikzpicture} = \bar V^{\mu_1\nu_1\,\mu_2\nu_2}(k_1, k_2),
\end{equation}
where
\begin{equation}
\begin{aligned}
    &\bar V^{\mu_1\nu_1\,\mu_2\nu_2}(k_1, k_2) = \int\mathrm{d}^dx\,\ee^{\ii q\cdot x} \\
    &\hspace{3em}\times\big(-2\ii\bar{\Phi}_{[\partial^0]}^{\mu_1\nu_1\,\mu_2\nu_2\,\gamma\delta}k_{1\gamma}k_{2\delta} + \bar{\Phi}_{[\partial^1]}^{\mu_1\nu_1\,\mu_2\nu_2\,\delta}k_{1\delta} + \bar{\Phi}_{[\partial^1]}^{\mu_2\nu_2\,\mu_1\nu_1\,\delta}k_{2\delta} + 2\ii\bar{\Phi}_{[\partial^2]}^{\mu_1\nu_1\,\mu_2\nu_2}\big).
\end{aligned}
\label{eq:2-point-se}\end{equation}
Note that the gray thick ``graviton" line attached to the massive body worldline is only introduced to denote all-order-in-$\kappa^2$ interactions with the background that are incorporated into this two-point vertex.\footnote{The contributions from the background metric map to interactions involving potential gravitons in a flat space expansion.} We emphasize that no properties of the Schwarzschild-Tangherlini metric were used in the derivation of this formula. Therefore, the vertex will take this form for any background geometry.

Let us now specialize to the Schwarzschild-Tangherlini metric in $d$ dimensions,
\begin{equation}
    \bar g_{\mu\nu}(x) = \Big(1 + \frac{\Lambda_d}{4\abs{x_\perp}^{d-3}}\Big)^\frac{4}{d-3}\,\eta_{\perp\mu\nu} + \frac{\big(1 - \frac{\Lambda_d}{4\abs{x_\perp}^{d-3}}\big)^2}{\big(1 + \frac{\Lambda_d}{4\abs{x_\perp}^{d-3}}\big)^2}\,\eta_{\parallel\mu\nu}.
\label{eq:bmetric}\end{equation}
Many choices of coordinate gauge are available; here we choose the isotropic one as was done in \cite{Kosmopoulos:2023bwc, Cheung:2023lnj}. In the interest of maintaining manifest Lorentz covariance, we have written the metric in an arbitrary asymptotic inertial frame by defining the restricted metrics,
\begin{subequations}
\begin{align}
    \eta_{\parallel\mu\nu} &\equiv v_\mu v_\nu, \\
    \eta_{\perp\mu\nu} &\equiv \eta_{\mu\nu} - v_\mu v_\nu,
\end{align}
\end{subequations}
which are projection operators onto the one-dimensional subspace collinear with the black hole velocity, $v^\mu$, satisfying $v^2 = 1$, and the $(d-1)$-dimensional subspace perpendicular to said velocity, respectively. When used on a vector such as $x^\mu$, we will adopt the self-explanatory notation,
\begin{equation}
    x_\perp^\mu \equiv \eta\indices*{^\mu_{\perp\nu}} x^\nu,
\end{equation}
as seen in eq.\,\eqref{eq:bmetric} in conjunction with $\abs{x_\perp} \equiv \sqrt{\abs{x_\perp^2}}$. We also define the dimensionally dependent scale,
\begin{equation}
    \Lambda_d \equiv \frac{\kappa^2M}{2(d-2)\Omega_{d-2}}, \qquad \text{with} \qquad \Omega_{d-2} \equiv \frac{2\pi^\frac{d-1}{2}}{\Gamma(\frac{d-1}{2})},
\label{eq:metric-scales}\end{equation}
where $\Omega_{d-2}$ is the surface area of the $(d-2)$-dimensional unit sphere. We note that the advantage of isotropic coordinates is twofold: Firstly, they can be easily written in terms of $\eta_{\mu\nu}$ and $v_\mu$ as in eq.\,\eqref{eq:bmetric}, making tensor manipulations simple. Secondly, when expanded, the powers of $G$ and $\abs{x_\perp}$ are correlated, so that an expression of order $G^n$ is an $(n-1)$-loop integral.

In this work, we wish to verify that we obtain the same result using the curved and flat space Feynman rules. Hence, we can expand the background metric in eq.\,\eqref{eq:bmetric},
\begin{equation}
    \bar g_{\mu\nu}(x) = \eta_{\mu\nu} - \frac{\Lambda_d}{\abs{x_\perp}^{d-3}}\Big(\eta_{\parallel\mu\nu} - \frac{\eta_{\perp\mu\nu}}{d-3}\Big) + \cdots,
\end{equation}
and insert it into eq.\,\eqref{eq:2-point-se}, after which the Fourier transform may be evaluated order by order in $\kappa^2$. Putting all of this together, the momentum space vertex factor can be expanded as
\begin{equation}
    \bar V^{\mu_1\nu_1\,\mu_2\nu_2}(k_1, k_2) = \sum_{i=1}^\infty\bar V_{(i)}^{\mu_1\nu_1\,\mu_2\nu_2}(k_1, k_2),
\end{equation}
depicted diagrammatically with
\begin{equation}
    \begin{tikzpicture}[baseline={(current bounding box.center)}]
        \coordinate (in) at (-1,0);
        \coordinate (out) at (1,0);
        \coordinate (x) at (0,0);
        \coordinate (gin) at (-1,-1.2);
        \coordinate (gout) at (1,-1.2);
        \coordinate (v) at (0,-1.2);
        \draw [dotted, thick] (in) -- (x);
        \draw [dotted, thick] (x) -- (out);
        \draw [photon2] (x) -- (v) node [midway, left, black] {$q\!\downarrow$};
        \draw [photon] (v) -- (gin) node [left, below=.5em] {$h_{\mu_1\nu_1}(k_1)$};
        \draw [photon] (v) -- (gout) node [right, below=.5em] {$h_{\mu_2\nu_2}(k_2)$};
        \draw [fill] (v) circle (.08);
        \node [draw,gray,thick,fill=white,circle,cross,minimum width=.3cm] at (x) {};
    \end{tikzpicture} = \sum_{i=1}^\infty\begin{tikzpicture}[baseline={(current bounding box.center)}]
        \coordinate (in) at (-1,0);
        \coordinate (out) at (1,0);
        \coordinate (x) at (0,0);
        \coordinate (gin) at (-1,-1.2);
        \coordinate (gout) at (1,-1.2);
        \coordinate (v) at (0,-1.2);
        \draw [dotted, thick] (in) -- (x);
        \draw [dotted, thick] (x) -- (out);
        \draw [photon2] (x) -- (v) node [midway, left, black] {$q\!\downarrow$};
        \draw [photon] (v) -- (gin) node [left, below=.5em] {$h_{\mu_1\nu_1}(k_1)$};
        \draw [photon] (v) -- (gout) node [right, below=.5em] {$h_{\mu_2\nu_2}(k_2)$};
        \draw [gray,fill=white,thick] (x) circle (.2) node {\tiny$i$};
        \draw [fill] (v) circle (.08);
    \end{tikzpicture}.
\label{eq:2-pt-expanded-diagram}\end{equation}
For our computation, we only need the first-order vertex,
\begin{equation}
    \begin{tikzpicture}[baseline={(current bounding box.center)}]
        \coordinate (in) at (-1,0);
        \coordinate (out) at (1,0);
        \coordinate (x) at (0,0);
        \coordinate (gin) at (-1,-1.2);
        \coordinate (gout) at (1,-1.2);
        \coordinate (v) at (0,-1.2);
        \draw [dotted, thick] (in) -- (x);
        \draw [dotted, thick] (x) -- (out);
        \draw [photon2] (x) -- (v) node [midway, left, black] {$q\!\downarrow$};
        \draw [photon] (v) -- (gin) node [left, below=.5em] {$h_{\mu_1\nu_1}(k_1)$};
        \draw [photon] (v) -- (gout) node [right, below=.5em] {$h_{\mu_2\nu_2}(k_2)$};
        \draw [gray,fill=white,thick] (x) circle (.2) node {\tiny$1$};
        \draw [fill] (v) circle (.08);
    \end{tikzpicture} = \ii\hat\delta(q\cdot v)\frac{\kappa^2M}{q^2}\bar N_{(1)}^{\mu_1\nu_1\,\mu_2\nu_2},
\label{eq:2-pt-kappa2}\end{equation}
where $\bar N_{(1)}^{\mu_1\nu_1\,\mu_2\nu_2}$ is a numerator containing the tensor structure of the vertex, and the second-order vertex,
\begin{equation}
    \begin{tikzpicture}[baseline={(current bounding box.center)}]
        \coordinate (in) at (-1,0);
        \coordinate (out) at (1,0);
        \coordinate (x) at (0,0);
        \coordinate (gin) at (-1,-1.2);
        \coordinate (gout) at (1,-1.2);
        \coordinate (v) at (0,-1.2);
        \draw [dotted, thick] (in) -- (x);
        \draw [dotted, thick] (x) -- (out);
        \draw [photon2] (x) -- (v) node [midway, left, black] {$q\!\downarrow$};
        \draw [photon] (v) -- (gin) node [left, below=.5em] {$h_{\mu_1\nu_1}(k_1)$};
        \draw [photon] (v) -- (gout) node [right, below=.5em] {$h_{\mu_2\nu_2}(k_2)$};
        \draw [gray,fill=white,thick] (x) circle (.2) node {\tiny$2$};
        \draw [fill] (v) circle (.08);
    \end{tikzpicture} = \ii\hat\delta(q\cdot v)(\kappa^2M)^2\int_\ell\frac{\hat\delta(\ell\cdot v)\bar N_{(2)}^{\mu_1\nu_1\,\mu_2\nu_2}}{\ell^2(q-\ell)^2}.
\label{eq:2-pt-kappa4}\end{equation}
To derive Feynman rules involving the worldline, it is helpful to decompose the background metric into the flat Minkowski metric and a remainder that accounts for the curvature,
\begin{equation}
    \bar g_{\mu\nu}(x) = \eta_{\mu\nu} + \bar\gamma_{\mu\nu}(x).
\end{equation}
Then, as in the case of the weak-field expansion, the linearity of the worldline action in the metric results in
\begin{equation}
    S_\text{wl}[\bar g + \kappa h,x] = S_\text{wl}[\eta,x] + \kappa S_\text{wl}[h,x] + S_\text{wl}[\bar\gamma,x].
\label{eq:wl-action-decomp-curved}\end{equation}
The first two terms in this expanded action are precisely those found in eq.\,\eqref{eq:wl-action-decomp}. The additional term, $S_\text{wl}[\bar\gamma,x]$, which includes the non-trivial, curved portion of the metric evaluated on the worldline, describes the self-force experienced by the body represented by the worldline. One might be concerned about the divergence that arises when $\bar\gamma_{\mu\nu}(x)$ is evaluated at the location of the black hole; however, as is shown in appendix \ref{app:scaleless} and also outlined in \cite{Cheung:2024byb, Kosmopoulos:2023bwc}, this part of the action is scaleless and vanishes in dimensional regularization. This leaves us with the same action as in the weak-field expansion,
\begin{equation}
    S_\text{wl}[\bar g + \kappa h,x] = S_\text{wl}[\eta + \kappa h,x],
\end{equation}
implying that the curved space Feynman rules for the worldline are identical to the ones derived from the weak-field expansion. The simplest of these rules is the source rule, eq.\,\eqref{eq:wl-source}, which we can rewrite in terms of the energy-momentum tensor of the background,
\begin{align}
    \bar T^{\mu\nu}(x) &= \frac{-2}{\bmetdens}\frac{\delta S_\text{wl}[\bar g,v\tau]}{\delta\bar g_{\mu\nu}(x)} \notag\\
    &= \frac{M}{\bmetdens}\int\mathrm{d}\tau\,\delta^{(d)}(x - v\tau)v^\mu v^\nu.
\end{align}
To wit,
\begin{equation}
    \kappa S_\text{wl}[h,x]\big\vert_{z^0} = -\frac{\kappa}{2}\int\mathrm{d}^dx\, \bmetdens \bar T^{\mu\nu}(x)h_{\mu\nu}(x).
\end{equation}
When this term is added to the linear part of the Einstein-Hilbert action from eq.\,\eqref{eq:lin-eh-term}, we obtain Einstein's equation for the background,
\begin{equation}
    S_\text{EH}[\bar g + \kappa h]\big\vert_{h^1} + \kappa S_\text{wl}[h,x]\big\vert_{z^0} = \frac{2}{\kappa}\int\mathrm{d}^dx\,\bmetdens\Big[\bar G^{\mu\nu}(x) -\frac{\kappa^2}{4}\bar T^{\mu\nu}(x)\Big]h_{\mu\nu}(x),
\end{equation}
which vanishes identically, as it should, when $\bar g_{\mu\nu}$ solves Einstein's equation (the Schwarzschild-Tangherlini metric being the solution in this case). Therefore, as the linear-in-$h$ terms of the action vanish, the curved space Feynman rules contain no vertex where the heavy body sources a graviton, killing this class of diagrams. The effect of these types of diagrams in flat space is reconfigured and encoded in the curved space Einstein-Hilbert graviton vertices through the presence of a non-flat metric. Diagrammatically speaking,
\begin{equation}
    \begin{tikzpicture}[baseline={(current bounding box.center)}]
        \coordinate (in) at (-1,0);
        \coordinate (out) at (1,0);
        \coordinate (x) at (0,0);
        \coordinate (g1) at (-1,-1.2);
        \coordinate (g2) at (-.7,-1.2-.7);
        \coordinate (gn) at (1,-1.2);
        \coordinate (v) at (0,-1.2);
        \draw [dotted, thick] (in) -- (x);
        \draw [dotted, thick] (x) -- (out);
        \draw [photon2] (x) -- (v);
        \draw [photon] (v) -- (g1);
        \draw [photon] (v) -- (g2);
        \draw [photon] (v) -- (gn);
        \draw [loosely dotted, thick] ([shift=(-110:.7)]0,-1.2) arc (-110:-25:.7);
        \node [gray,draw,thick,fill=white,circle,cross,minimum width=.3cm] at (x) {};
        \draw [fill] (v) circle (.08);
    \end{tikzpicture} = \begin{tikzpicture}[baseline={(current bounding box.center)}]
        \coordinate (in) at (-1,0);
        \coordinate (out) at (1,0);
        \coordinate (x) at (0,0);
        \coordinate (g1) at (-1,-1.2);
        \coordinate (g2) at (-.7,-1.2-.7);
        \coordinate (gn) at (1,-1.2);
        \coordinate (v) at (0,-1.2);
        \draw [dotted, thick] (in) -- (x);
        \draw [photon] (x) -- (v);
        \draw [dotted, thick] (x) -- (out);
        \draw [photon] (v) -- (g1);
        \draw [photon] (v) -- (g2);
        \draw [photon] (v) -- (gn);
        \draw [loosely dotted, thick] ([shift=(-110:.7)]0,-1.2) arc (-110:-25:.7);
        \draw [fill] (v) circle (.04);
        \draw [fill] (x) circle (.08);
    \end{tikzpicture} + \frac{1}{2!}\hspace{2pt}\begin{tikzpicture}[baseline={(current bounding box.center)}]
        \coordinate (in) at (-1,0);
        \coordinate (out) at (1,0);
        \coordinate (x1) at (-.6,0);
        \coordinate (x2) at (.6,0);
        \coordinate (g1) at (-1,-1.2);
        \coordinate (g2) at (-.7,-1.2-.7);
        \coordinate (gn) at (1,-1.2);
        \coordinate (v1) at (0,-.5);
        \coordinate (v2) at (0,-1.2);
        \draw [dotted, thick] (in) -- (out);
        \draw [photon] (x1) -- (v1);
        \draw [photon] (x2) -- (v1);
        \draw [photon] (v1) -- (v2);
        \draw [photon] (v2) -- (g1);
        \draw [photon] (v2) -- (g2);
        \draw [photon] (v2) -- (gn);
        \draw [loosely dotted, thick] ([shift=(-110:.7)]0,-1.2) arc (-110:-25:.7);
        \draw [fill] (v1) circle (.04);
        \draw [fill] (v2) circle (.04);
        \draw [fill] (x1) circle (.08);
        \draw [fill] (x2) circle (.08);
    \end{tikzpicture} + \frac{1}{2!}\hspace{2pt}\begin{tikzpicture}[baseline={(current bounding box.center)}]
        \coordinate (in) at (-1,0);
        \coordinate (out) at (1,0);
        \coordinate (x1) at (-.6,0);
        \coordinate (x2) at (.6,0);
        \coordinate (g1) at (-1,-1.2);
        \coordinate (g2) at (-.7,-1.2-.7);
        \coordinate (gn) at (1,-1.2);
        \coordinate (v) at (0,-1.2);
        \draw [dotted, thick] (in) -- (out);
        \draw [photon] (x1) -- (v);
        \draw [photon] (x2) -- (v);
        \draw [photon] (v) -- (g1);
        \draw [photon] (v) -- (g2);
        \draw [photon] (v) -- (gn);
        \draw [loosely dotted, thick] ([shift=(-110:.7)]0,-1.2) arc (-110:-25:.7);
        \draw [fill] (v) circle (.04);
        \draw [fill] (x1) circle (.08);
        \draw [fill] (x2) circle (.08);
    \end{tikzpicture} + \cdots,
\label{eq:se-wfe-correspondence}\end{equation}
where the left-hand side depicts a curved space $n$-point graviton vertex that contributes to an infinite order in $\kappa^2$ and the right-hand side shows the equivalent set of flat space diagrams. We note that equality between the two sides is modulo gauge-dependent terms. Notice that the second and third diagrams on the right-hand side contain what is effectively a one-loop integral. Looking at the second-order graviton vertex, eq.\,\eqref{eq:2-pt-kappa4}, we can see how such loop integrals emerge from curved space rules. An attractive feature of this resummation is that the curved space graviton vertex, if connected to external momenta, will only ever contain two loop momenta in the numerator. In contrast, the numerator of a $\kappa^{2n}$ order weak-field diagram can contain up to $2(n-1)$ momenta, making this effective tensor reduction of the integrand quite considerable at higher post-Minkowskian orders. It is worth noting that the absence of a source rule in the curved space expansion means that it is not possible to draw diagrams containing self-energy pieces such as
\begin{equation}
    \begin{tikzpicture}[baseline={(current bounding box.center)}]
        \coordinate (in) at (-2,0);
        \coordinate (out) at (2,0);
        \coordinate (x1) at (-1.3,0);
        \coordinate (x2) at (0,0);
        \coordinate (x3) at (1.3,0);
        \coordinate (gout) at (1.3,-1);
        \draw [dotted, thick] (in) -- (out);
        \draw [zUndirected] (x2) -- (x3);
        \draw [photon] (x1) to [bend right=80] (x2);
        \draw [photon] (x3) -- (gout);
        \draw [fill] (x1) circle (.08);
        \draw [fill] (x2) circle (.08);
        \draw [fill] (x3) circle (.08);
    \end{tikzpicture}.
\end{equation}
In the weak-field expansion, one typically encounters them and must exclude them manually.
An alternative to the worldline rules discussed above, involves integration of the deflection out of the action, giving rise to effective graviton interactions that encode the recoil of the massive source \cite{Cheung:2023lnj,Cheung:2024byb}. To quadratic order in the graviton field, we get
\begin{equation}
S_{\rm recoil} =   - \frac{\kappa^2  M}{2} \int d\tau \,   v^{\alpha} v^{\beta} \delta \Gamma^\mu_{\;\;\alpha\beta}(v \tau) \frac{1}{\partial_\tau^2} v^{\gamma}  v^{\delta} \delta \Gamma_{\mu\gamma\delta}(v \tau) \,,
\end{equation}
where $\delta \Gamma^\rho_{\mu\nu}  \equiv \Gamma^\rho_{\mu\nu}-\bar\Gamma^\rho_{\mu\nu} = \tfrac{\kappa}{2} \bar g ^{\rho \sigma}(\bar\nabla_\mu h_{\sigma \nu}+\bar\nabla_\nu h_{\sigma \mu}-\bar\nabla_\sigma h_{\mu \nu} )$ is a gauge-invariant difference of connections with respect to the background metric. We are then left with a single explicit dynamical field, $h_{\mu \nu}$, in our theory, and an additional graviton vertex of the form,
\begin{align}
\label{recoil}
\begin{tikzpicture}[baseline={(current bounding box.center)}]
        \coordinate (in) at (0,0);
        \coordinate (out) at (1.2,0);
        \coordinate (x1) at (0.4,0);
        \coordinate (x2) at (0.8,0);
        \coordinate (gin) at (0,-.8);
        \coordinate (gout) at (1.2,-.8);
        \draw [dotted, thick] (in) -- (x1);
        \draw [dotted, thick] (x2) -- (out);
        \draw [zUndirected] (x1) -- (x2);
        \draw [photon] (x1) -- (gin) node [left=.6em, below] {$h_{\mu_1\nu_1}(k_1)$};
        \draw [photon] (x2) -- (gout) node [right=.6em, below] {$h_{\mu_2\nu_2}(k_2)$};
        \draw [fill] (x1) circle (.08);
        \draw [fill] (x2) circle (.08);
    \end{tikzpicture} &= \frac{\ii \kappa^2 M \hat\delta((k_1+k_2)\cdot v)}{4(k_1 \cdot v)(k_2 \cdot v)}\\
    &\times (2 (k_1 \cdot v) v^{(\mu_1} \eta^{\nu_1) \rho}-v^{\mu_1} v^{\nu_1} k^\rho_1)(2 (k_2 \cdot v) v^{(\mu_2} \eta^{\nu_2)}_{\rho}-v^{\mu_1} v^{\nu_1} k_{2\rho}).\notag
\end{align}

\subsection{The classical Compton amplitude}
\label{sec:comp_amp}
In the curved space setup, the diagrammatic expansion of the Compton amplitude takes a very structured form,
\begin{equation}\label{eq:ComptonGen}
\begin{aligned}
    \begin{tikzpicture}[baseline={(current bounding box.center)}]
        \coordinate (in) at (-1,0);
        \coordinate (out) at (1,0);
        \coordinate (gin) at (-1,-.93);
        \coordinate (gout) at (1,-.93);
        \draw [dotted, thick] (in) -- (out);
        \draw [photon] (gin) -- (-.25,-.45);
        \draw [photon] (gout) -- (.25,-.45);
        \fill [darkgray] (0,-.25) circle (.4);
    \end{tikzpicture} &= \begin{tikzpicture}[baseline={(current bounding box.center)}]
        \coordinate (in) at (0,0);
        \coordinate (out) at (1.2,0);
        \coordinate (x) at (.6,0);
        \coordinate (gin) at (0,-.8);
        \coordinate (gout) at (1.2,-.8);
        \coordinate (v) at (.6,-.8);
        \draw [dotted, thick] (in) -- (x);
        \draw [dotted, thick] (x) -- (out);
        \draw [photon2] (x) -- (v);
        \draw [photon] (v) -- (gin);
        \draw [photon] (v) -- (gout);
        \draw [fill] (v) circle (.08);
        \node [gray,draw,thick,fill=white,circle,cross,minimum width=.3cm] at (x) {};
    \end{tikzpicture} + \begin{tikzpicture}[baseline={(current bounding box.center)}]
        \coordinate (in) at (0,0);
        \coordinate (out) at (2.0,0);
        \coordinate (x1) at (0.6,0);
        \coordinate (x2) at (1.4,0);
        \coordinate (gin) at (0,-.8);
        \coordinate (gout) at (2.0,-.8);
        \coordinate (v1) at (0.6,-.8);
        \coordinate (v2) at (1.4,-.8);
        \draw [dotted, thick] (in) -- (out);
        \draw [photon2] (x1) -- (v1);
        \draw [photon2] (x2) -- (v2);
        \draw [photon] (v1) -- (gin);
        \draw [photon] (v1) -- (v2);
        \draw [photon] (v2) -- (gout);
        \draw [fill] (v1) circle (.08);
        \node [gray,draw,thick,fill=white,circle,cross,minimum width=.3cm] at (x1) {};
        \draw [fill] (v2) circle (.08);
        \node [gray,draw,thick,fill=white,circle,cross,minimum width=.3cm] at (x2) {};
    \end{tikzpicture} + \cdots \\
    &+ \begin{tikzpicture}[baseline={(current bounding box.center)}]
        \coordinate (in) at (0,0);
        \coordinate (out) at (1.2,0);
        \coordinate (x1) at (0.4,0);
        \coordinate (x2) at (0.8,0);
        \coordinate (gin) at (0,-.8);
        \coordinate (gout) at (1.2,-.8);
        \draw [dotted, thick] (in) -- (x1);
        \draw [dotted, thick] (x2) -- (out);
        \draw [zUndirected] (x1) -- (x2);
        \draw [photon] (x1) -- (gin);
        \draw [photon] (x2) -- (gout);
        \draw [fill] (x1) circle (.08);
        \draw [fill] (x2) circle (.08);
    \end{tikzpicture} + \Bigg(\begin{tikzpicture}[baseline={(current bounding box.center)}]
        \coordinate (in) at (0,0);
        \coordinate (out) at (2.0,0);
        \coordinate (x1) at (0.4,0);
        \coordinate (x2) at (0.8,0);
        \coordinate (x3) at (1.4,0);
        \coordinate (v2) at (1.4,-.8);
        \coordinate (gin) at (0,-.8);
        \coordinate (gout) at (2.0,-.8);
        \draw [dotted, thick] (in) -- (x1);
        \draw [dotted, thick] (x2) -- (out);
        \draw [zUndirected] (x1) -- (x2);
        \draw [photon] (x1) -- (gin);
        \draw [photon] (x2) -- (v2);
        \draw [photon2] (x3) -- (v2);
        \draw [photon] (v2) -- (gout);
        \draw [fill] (x1) circle (.08);
        \draw [fill] (x2) circle (.08);
        \draw [fill] (v2) circle (.08);
        \node [gray,draw,thick,fill=white,circle,cross,minimum width=.3cm] at (x3) {};
    \end{tikzpicture} + \text{perms}\Bigg) + \cdots \\
    &+ \begin{tikzpicture}[baseline={(current bounding box.center)}]
        \coordinate (in) at (0,0);
        \coordinate (out) at (2.2,0);
        \coordinate (x1) at (0.4,0);
        \coordinate (x2) at (0.8,0);
        \coordinate (x3) at (1.4,0);
        \coordinate (x4) at (1.8,0);
        \coordinate (gin) at (0,-.8);
        \coordinate (gout) at (2.2,-.8);
        \draw [dotted, thick] (in) -- (x1);
        \draw [dotted, thick] (x2) -- (out);
        \draw [zUndirected] (x1) -- (x2);
        \draw [photon] (x1) -- (gin);
        \draw [photon] (x2) to [bend right=60] (x3);
        \draw [zUndirected] (x3) -- (x4);
        \draw [photon] (x4) -- (gout);
        \draw [fill] (x1) circle (.08);
        \draw [fill] (x2) circle (.08);
        \draw [fill] (x3) circle (.08);
        \draw [fill] (x4) circle (.08);
    \end{tikzpicture} + \Bigg(\begin{tikzpicture}[baseline={(current bounding box.center)}]
        \coordinate (in) at (0,0);
        \coordinate (out) at (3.0,0);
        \coordinate (x1) at (0.4,0);
        \coordinate (x2) at (0.8,0);
        \coordinate (x3) at (1.4,0);
        \coordinate (x4) at (1.8,0);
        \coordinate (gin) at (0,-.8);
        \coordinate (v) at (2.4,-.8);
        \coordinate (x5) at (2.4,0);
        \coordinate (gout) at (3.0,-.8);
        \draw [dotted, thick] (in) -- (x1);
        \draw [dotted, thick] (x2) -- (out);
        \draw [zUndirected] (x1) -- (x2);
        \draw [photon] (x1) -- (gin);
        \draw [photon] (x2) to [bend right=60] (x3);
        \draw [zUndirected] (x3) -- (x4);
        \draw [photon] (x4) -- (v);
        \draw [photon2] (x5) -- (v);
        \draw [photon] (v) -- (gout);
        \draw [fill] (x1) circle (.08);
        \draw [fill] (x2) circle (.08);
        \draw [fill] (x3) circle (.08);
        \draw [fill] (x4) circle (.08);
        \draw [fill] (v) circle (.08);
        \node [gray,draw,thick,fill=white,circle,cross,minimum width=.3cm] at (x5) {};
    \end{tikzpicture} + \text{perms}\Bigg) + \cdots \\
    &+ \cdots.
\end{aligned}
\end{equation}
The gray blob represents the complete Compton amplitude, and `perms' denotes all possible permutations of background injections and deflections. If deflections of the massive worldline were turned off, only the top line would contribute. This would represent the scattering of a graviton off a fixed Schwarzschild-Tangherlini background. In the weak-field expansion, the contributions are much more involved,
\begin{equation}\label{eq:flat2pm}
\begin{aligned}
    \begin{tikzpicture}[baseline={(current bounding box.center)}]
        \coordinate (in) at (-1,0);
        \coordinate (out) at (1,0);
        \coordinate (gin) at (-1,-.93);
        \coordinate (gout) at (1,-.93);
        \draw [dotted, thick] (in) -- (out);
        \draw [photon] (gin) -- (-.25,-.45);
        \draw [photon] (gout) -- (.25,-.45);
        \fill [darkgray] (0,-.25) circle (.4);
    \end{tikzpicture} &= \begin{tikzpicture}[baseline={(current bounding box.center)}]
        \coordinate (in) at (0,0);
        \coordinate (out) at (1.2,0);
        \coordinate (x) at (.6,0);
        \coordinate (gin) at (0,-.8);
        \coordinate (gout) at (1.2,-.8);
        \coordinate (v) at (.6,-.8);
        \draw [dotted, thick] (in) -- (x);
        \draw [dotted, thick] (x) -- (out);
        \draw [photon] (x) -- (v);
        \draw [photon] (v) -- (gin);
        \draw [photon] (v) -- (gout);
        \draw [fill] (v) circle (.04);
        \draw [fill] (x) circle (.08);
    \end{tikzpicture} + \begin{tikzpicture}[baseline={(current bounding box.center)}]
        \coordinate (in) at (0,0);
        \coordinate (out) at (1.2,0);
        \coordinate (x1) at (0.4,0);
        \coordinate (x2) at (0.8,0);
        \coordinate (gin) at (0,-.8);
        \coordinate (gout) at (1.2,-.8);
        \draw [dotted, thick] (in) -- (x1);
        \draw [dotted, thick] (x2) -- (out);
        \draw [zUndirected] (x1) -- (x2);
        \draw [photon] (x1) -- (gin);
        \draw [photon] (x2) -- (gout);
        \draw [fill] (x1) circle (.08);
        \draw [fill] (x2) circle (.08);
    \end{tikzpicture} \\
    &+ \begin{tikzpicture}[baseline={(current bounding box.center)}]
        \coordinate (in) at (0,0);
        \coordinate (out) at (2.2,0);
        \coordinate (x1) at (0.4,0);
        \coordinate (x2) at (0.8,0);
        \coordinate (x3) at (1.4,0);
        \coordinate (x4) at (1.8,0);
        \coordinate (gin) at (0,-.8);
        \coordinate (gout) at (2.2,-.8);
        \draw [dotted, thick] (in) -- (x1);
        \draw [dotted, thick] (x2) -- (out);
        \draw [zUndirected] (x1) -- (x2);
        \draw [photon] (x1) -- (gin);
        \draw [photon] (x2) to [bend right=60] (x3);
        \draw [zUndirected] (x3) -- (x4);
        \draw [photon] (x4) -- (gout);
        \draw [fill] (x1) circle (.08);
        \draw [fill] (x2) circle (.08);
        \draw [fill] (x3) circle (.08);
        \draw [fill] (x4) circle (.08);
    \end{tikzpicture} + \Bigg(\begin{tikzpicture}[baseline={(current bounding box.center)}]
        \coordinate (in) at (0,0);
        \coordinate (out) at (2.0,0);
        \coordinate (x1) at (0.4,0);
        \coordinate (x2) at (0.8,0);
        \coordinate (x3) at (1.5,0);
        \coordinate (v2) at (1.5,-.8);
        \coordinate (gin) at (0,-.8);
        \coordinate (gout) at (2.0,-.8);
        \draw [dotted, thick] (in) -- (x1);
        \draw [dotted, thick] (x2) -- (out);
        \draw [zUndirected] (x1) -- (x2);
        \draw [photon] (x1) -- (gin);
        \draw [photon] (v2) -- (x2);
        \draw [photon] (v2) -- (x3);
        \draw [photon] (v2) -- (gout);
        \draw [fill] (x1) circle (.08);
        \draw [fill] (x2) circle (.08);
        \draw [fill] (v2) circle (.04);
        \draw [fill] (x3) circle (.08);
    \end{tikzpicture} + \text{perms}\Bigg) \\
    &+ \frac12\Bigg(\begin{tikzpicture}[baseline={(current bounding box.center)}]
        \coordinate (in) at (0,0);
        \coordinate (out) at (1.8,0);
        \coordinate (x1) at (0.6,0);
        \coordinate (x2) at (1.2,0);
        \coordinate (gin) at (0,-.8);
        \coordinate (gout) at (1.8,-.8);
        \coordinate (v1) at (.9,-.4);
        \coordinate (v2) at (.9,-.8);
        \draw [dotted, thick] (in) -- (out);
        \draw [photon] (x1) -- (v1);
        \draw [photon] (x2) -- (v1);
        \draw [photon] (v1) -- (v2);
        \draw [photon] (v2) -- (gin);
        \draw [photon] (v2) -- (gout);
        \draw [fill] (v1) circle (.04);
        \draw [fill] (v2) circle (.04);
        \draw [fill] (x1) circle (.08);
        \draw [fill] (x2) circle (.08);
    \end{tikzpicture} + \begin{tikzpicture}[baseline={(current bounding box.center)}]
        \coordinate (in) at (0,0);
        \coordinate (out) at (1.8,0);
        \coordinate (x1) at (0.6,0);
        \coordinate (x2) at (1.2,0);
        \coordinate (gin) at (0,-.8);
        \coordinate (gout) at (1.8,-.8);
        \coordinate (v) at (.9,-.8);
        \draw [dotted, thick] (in) -- (out);
        \draw [photon] (x1) -- (v);
        \draw [photon] (x2) -- (v);
        \draw [photon] (v) -- (gin);
        \draw [photon] (v) -- (gout);
        \draw [fill] (v) circle (.04);
        \draw [fill] (x1) circle (.08);
        \draw [fill] (x2) circle (.08);
    \end{tikzpicture}\Bigg) + \begin{tikzpicture}[baseline={(current bounding box.center)}]
        \coordinate (in) at (0,0);
        \coordinate (out) at (1.8,0);
        \coordinate (x1) at (0.6,0);
        \coordinate (x2) at (1.2,0);
        \coordinate (gin) at (0,-.8);
        \coordinate (gout) at (1.8,-.8);
        \coordinate (v1) at (0.6,-.8);
        \coordinate (v2) at (1.2,-.8);
        \draw [dotted, thick] (in) -- (out);
        \draw [photon] (x1) -- (v1);
        \draw [photon] (x2) -- (v2);
        \draw [photon] (v1) -- (gin);
        \draw [photon] (v1) -- (v2);
        \draw [photon] (v2) -- (gout);
        \draw [fill] (v1) circle (.04);
        \draw [fill] (x1) circle (.08);
        \draw [fill] (v2) circle (.04);
        \draw [fill] (x2) circle (.08);
    \end{tikzpicture} \\
    &+ \cdots.
\end{aligned}
\end{equation}
Here, post-Minkowskian order counting is straightforward, as each diagram comes with a well-defined power of $\kappa^2$. We display the Compton amplitude up to second post-Minkowskian order in the equation above. The symmetry factor of two comes from identical sources connecting to the same vertex. However, the presence of arbitrary-multiplicity vertices makes the expansion quite complicated.

\section{First post-Minkowskian order Compton amplitude}\label{sec:1pm}

In this section, we discuss the computation of the gravitational Compton amplitude at first post-Minkowskian order. We first compute using the curved space expansion as described in \cite{Kosmopoulos:2023bwc}, then confirm that the same result is obtained with the weak-field rules. At this low order, both computations involve two diagrams of similar complexity.
The first diagram involves the vertex from eq.\,\eqref{eq:2-pt-kappa2}, which we contract with external graviton polarization tensors satisfying
\begin{equation}
    \varepsilon_{i\mu\nu} = \varepsilon_{i\nu\mu}, \qquad p_i^\mu\varepsilon_{i\mu\nu} = 0, \qquad \varepsilon\indices{_{i\mu}^\mu} = 0.
\end{equation}
The second diagram involves a deflection mode traveling on the worldline,
\begin{equation}
\begin{aligned}
    &\begin{tikzpicture}[baseline={(current bounding box.center)}]
        \coordinate (in) at (-1.3,0);
        \coordinate (out) at (1.3,0);
        \coordinate (x1) at (-.6,0);
        \coordinate (x2) at (.6,0);
        \coordinate (gin) at (-.6,-1);
        \coordinate (gout) at (.6,-1);
        \draw [dotted, thick] (in) -- (x1);
        \draw [dotted, thick] (x2) -- (out);
        \draw [zUndirected] (x1) -- (x2) node [midway, above] {$\stackrel{\displaystyle\omega}{\rightarrow}$};
        \draw [photon] (x1) -- (gin) node [below] {$p_1$} node [midway, left=.3em] {$\uparrow$};
        \draw [photon] (x2) -- (gout) node [below] {$p_2$} node [midway, right=.3em] {$\downarrow$};;
        \draw [fill] (x1) circle (.08);
        \draw [fill] (x2) circle (.08);
    \end{tikzpicture} \\
    &= \ii\hat\delta(q\cdot v)\frac{\kappa^2M}{4\omega^2}\varepsilon_{1\mu_1\nu_1}\varepsilon_{2\mu_2\nu_2}(v^{\mu_1}v^{\nu_1}p_1^\rho - 2\omega v\indices{^{(\mu_1}}\eta^{\nu_1)\rho})(v^{\mu_2}v^{\nu_2}p_{2\rho} - 2\omega v\indices{^{(\mu_2}}\delta\indices*{^{\nu_2)}_\rho}).
\end{aligned}
\label{eq:1pm-deflection}\end{equation}
Here, we have integrated out the deflection energy using the delta function from one of the vertices and fixed the energy flowing through the worldline to $\omega = p_1\cdot v$, the energy of the graviton. The resulting expression is equivalent to the contraction of external graviton polarization tensors with the recoil vertex in eq.\,\eqref{recoil}. Summing the contributions from both diagrams, we note that all dependence on $d$ cancels automatically and the result is independent of the dimension. If we factorize the polarization tensors,
\begin{equation}
    \varepsilon_{i\mu\nu} \equiv \varepsilon_{i\mu}\varepsilon_{i\nu},
\end{equation}
we can recover the familiar gauge-invariant double copy form \cite{Brandhuber:2021kpo}  of the classical first post-Minkowskian order Compton amplitude,
\begin{equation}
    \ii\hat\delta(q\cdot v)\mathcal{M}^\text{1PM}(p_1, p_2) = -\ii\hat\delta(q\cdot v)\frac{\kappa^2M}{2}\frac{(v\cdot f_1\cdot f_2\cdot v)^2}{q^2\omega^2},
\label{eq:1pm-compton-flat}\end{equation}
expressed in terms of the momentum space field strength tensor of the incoming and outgoing gravitons,
\begin{equation}
    f_{i\mu\nu} \equiv p_{i\mu}\varepsilon_{i\nu} - p_{i\nu}\varepsilon_{i\mu},
\end{equation}
using the notation, $(V\cdot f_i)^\mu = V^\nu f\indices{_{i\nu}^\mu}$. As noted in \cite{Kosmopoulos:2023bwc}, the normalization of the amplitude in eq.\,\eqref{eq:1pm-compton-flat} differs by a factor of $2M$ from the conventional result, $\widetilde{\mathcal{M}}^\text{1PM}(p_1,p_2)$. This arises purely from different conventions as seen from
\begin{equation}
    \ii\hat\delta(2q\cdot Mv)\widetilde{\mathcal{M}}^\text{1PM}(p_1,p_2) = \ii\hat\delta(q\cdot v)\mathcal{M}^\text{1PM}(p_1, p_2).
\end{equation}
Using the weak-field rules instead, we get
\begin{equation}
    \begin{tikzpicture}[baseline={(current bounding box.center)}]
        \coordinate (in) at (-1,0);
        \coordinate (out) at (1,0);
        \coordinate (x) at (0,0);
        \coordinate (gin) at (-1,-1.2);
        \coordinate (gout) at (1,-1.2);
        \coordinate (v) at (0,-1.2);
        \draw [dotted, thick] (in) -- (x);
        \draw [dotted, thick] (x) -- (out);
        \draw [photon] (x) -- (v) node [midway, left] {$q\!\downarrow$};
        \draw [photon] (v) -- (gin) node [left] {$p_1$} node [midway, below=.3em] {$\rightarrow$};
        \draw [photon] (v) -- (gout) node [right] {$p_2$} node [midway, below=.3em] {$\rightarrow$};;
        \draw [fill] (v) circle (.04);
        \draw [fill] (x) circle (.08);
    \end{tikzpicture} = \ii\hat\delta(q\cdot v)\frac{\kappa^2M}{2}\varepsilon_{1\mu_1\nu_1}\varepsilon_{2\mu_2\nu_2}V^{\mu_1\nu_1\,\rho\sigma\,\mu_2\nu_2}(-p_1,-q, p_2)\frac{P_{\rho\sigma\,\gamma\delta}}{q^2}v^\gamma v^\delta,
\label{eq:1pm-3pt}\end{equation}
where the integral over the momentum emitted from the worldline fixes $q = p_2 - p_1$ due to the delta function from the three-point vertex (cf. eq.\,\eqref{eq:flat2pm}). Since the deflection diagram is the same in both expansions, we find the Compton amplitude from the weak-field calculation,
\begin{equation}
    \ii\hat\delta(q\cdot v)\mathcal{M}^\text{1PM}(p_1, p_2) = \eqref{eq:1pm-3pt} + \eqref{eq:1pm-deflection},
\end{equation}
which, as indicated, coincides with the result obtained from the expansion in curved space. This is an example of the correspondence depicted in eq.\,\eqref{eq:se-wfe-correspondence} at order $\kappa^2$.

\section{Second post-Minkowskian order Compton amplitude}\label{sec:2pm}
Let us now discuss the topic that is the focus of this paper, the classical Compton amplitude at second post-Minkowskian order. This is the lowest order at which the resummation of potential graviton diagrams becomes apparent as two flat space diagrams combine into one in curved space,
\begin{equation}
    \begin{tikzpicture}[baseline={(current bounding box.center)}]
        \coordinate (in) at (-1,0);
        \coordinate (out) at (1,0);
        \coordinate (x1) at (-.6,0);
        \coordinate (x2) at (.6,0);
        \coordinate (g1) at (-1,-1.2);
        \coordinate (g2) at (1,-1.2);
        \coordinate (v1) at (0,-.5);
        \coordinate (v2) at (0,-1.2);
        \draw [dotted, thick] (in) -- (out);
        \draw [photon] (x1) -- (v1);
        \draw [photon] (x2) -- (v1);
        \draw [photon] (v1) -- (v2);
        \draw [photon] (v2) -- (g1);
        \draw [photon] (v2) -- (g2);
        \draw [fill] (v1) circle (.04);
        \draw [fill] (v2) circle (.04);
        \draw [fill] (x1) circle (.08);
        \draw [fill] (x2) circle (.08);
    \end{tikzpicture} \quad\text{and}\quad \begin{tikzpicture}[baseline={(current bounding box.center)}]
        \coordinate (in) at (-1,0);
        \coordinate (out) at (1,0);
        \coordinate (x1) at (-.6,0);
        \coordinate (x2) at (.6,0);
        \coordinate (g1) at (-1,-1.2);
        \coordinate (g2) at (1,-1.2);
        \coordinate (v) at (0,-1.2);
        \draw [dotted, thick] (in) -- (out);
        \draw [photon] (x1) -- (v);
        \draw [photon] (x2) -- (v);
        \draw [photon] (v) -- (g1);
        \draw [photon] (v) -- (g2);
        \draw [fill] (v) circle (.04);
        \draw [fill] (x1) circle (.08);
        \draw [fill] (x2) circle (.08);
    \end{tikzpicture} \qquad\rightarrow\qquad \begin{tikzpicture}[baseline={(current bounding box.center)}]
        \coordinate (in) at (-1,0);
        \coordinate (out) at (1,0);
        \coordinate (x) at (0,0);
        \coordinate (gin) at (-1,-1.2);
        \coordinate (gout) at (1,-1.2);
        \coordinate (v) at (0,-1.2);
        \draw [dotted, thick] (in) -- (x);
        \draw [dotted, thick] (x) -- (out);
        \draw [photon2] (x) -- (v);
        \draw [photon] (v) -- (gin);
        \draw [photon] (v) -- (gout);
        \draw [gray,fill=white,thick] (x) circle (.2) node {\tiny$2$};
        \draw [fill] (v) circle (.08);
    \end{tikzpicture}.
\end{equation}
As with the amplitude at first post-Minkowskian order, we first obtain the integrand using the curved space expansion and then confirm that the weak-field rules lead to the same result in flat space.

\subsection{Diagrammatic expansion}
To obtain the second post-Minkowskian order piece of eq.\,\eqref{eq:ComptonGen}, we use the expanded vertex rules in eqs. \eqref{eq:2-pt-kappa2} and \eqref{eq:2-pt-kappa4}. We get five diagrams, labeled (i)--(v), through this procedure. Diagram (i) is of a radiation-reaction type with the structure,
\begin{equation}
    \begin{tikzpicture}[baseline={(current bounding box.center)}]
        \coordinate (in) at (-2,0);
        \coordinate (out) at (2,0);
        \coordinate (x1) at (-1.5,0);
        \coordinate (x2) at (-.5,0);
        \coordinate (x3) at (.5,0);
        \coordinate (x4) at (1.5,0);
        \coordinate (gin) at (-1.5,-1);
        \coordinate (gout) at (1.5,-1);
        \draw [dotted, thick] (in) -- (x1);
        \draw [dotted, thick] (x2) -- (x3);
        \draw [dotted, thick] (x4) -- (out);
        \draw [zUndirected] (x1) -- (x2) node [midway, above] {$\stackrel{\displaystyle\omega}{\rightarrow}$};
        \draw [zUndirected] (x3) -- (x4) node [midway, above] {$\stackrel{\displaystyle\omega}{\rightarrow}$};
        \draw [photon] (x1) -- (gin) node [below] {$p_1$} node [midway, left=.3em] {$\uparrow$};
        \draw [photon] (x2) to [bend right=80] (x3) node [midway, below=.8em] {$\stackrel{\displaystyle\rightarrow}{p_1 + \ell}$};
        \draw [photon] (x4) -- (gout) node [below=-.3em] {$p_2$} node [midway, right=.3em] {$\downarrow$};
        \draw [fill] (x1) circle (.08);
        \draw [fill] (x2) circle (.08);
        \draw [fill] (x3) circle (.08);
        \draw [fill] (x4) circle (.08);
    \end{tikzpicture} = \ii\hat\delta(q\cdot v)\frac{\kappa^4M^2}{\omega^4}\int_\ell\frac{\hat\delta(\ell\cdot v)N_\text{i}(\varepsilon_1,\varepsilon_2,\ell)}{(p_1 + \ell)^2}.
\end{equation}
We have contracted the graviton polarization vectors and collected the numerator in $N_\text{i}$. Diagram (ii) involves the Feynman rule from eq.\,\eqref{eq:2-pt-kappa2},
\begin{equation}
    \begin{tikzpicture}[baseline={(current bounding box.center)}]
        \coordinate (in) at (-1.5,0);
        \coordinate (out) at (2.5,0);
        \coordinate (x1) at (-1,0);
        \coordinate (y1) at (-.25,-.8);
        \coordinate (x2) at (.5,0);
        \coordinate (x3) at (2,0);
        \coordinate (gin) at (-.25,-1.4);
        \coordinate (gout) at (2,-1.4);
        \draw [dotted, thick] (in) -- (x2);
        \draw [dotted, thick] (x4) -- (out);
        \draw [zUndirected] (x2) -- (x3) node [midway, above] {$\stackrel{\displaystyle\omega}{\rightarrow}$};
        \draw [photon] (y1) -- (gin) node [below] {$p_1$} node [midway, left=.3em] {$\uparrow$};
        \draw [photon2] (x1) -- (y1) node [midway, left, black] {$\ell\searrow$};
        \draw [photon] (y1) -- (x2) node [midway, right] {$\nearrow p_1+\ell$};
        \draw [photon] (x3) -- (gout) node [below] {$p_2$} node [midway, right=.3em] {$\downarrow$};
        \draw [fill] (x2) circle (.08);
        \draw [fill] (x3) circle (.08);
        \draw [gray,fill=white,thick] (x1) circle (.2) node {\tiny$1$};
        \draw [fill] (y1) circle (.08);
    \end{tikzpicture} = \ii\hat\delta(q\cdot v)\frac{\kappa^4M^2}{\omega^2}\int_\ell\frac{\hat\delta(\ell\cdot v)N_\text{ii}(\varepsilon_1,\varepsilon_2,\ell)}{\ell^2(p_1 + \ell)^2}.
\end{equation}
A mirrored version of diagram (ii), labeled diagram (iii), is also included. It can be easily obtained from (ii) by swapping $(p_1,\varepsilon_1)$ with $(p_2,\varepsilon_2)$ and changing the sign of $\omega$. Next, we come to diagram (iv) containing the second-order piece of the curved space vertex from eq.\,\eqref{eq:2-pt-kappa4},
\begin{equation}
    \begin{tikzpicture}[baseline={(current bounding box.center)}]
        \coordinate (in) at (-1,0);
        \coordinate (out) at (1,0);
        \coordinate (x) at (0,0);
        \coordinate (gin) at (-.65,-1.6);
        \coordinate (gout) at (.65,-1.6);
        \coordinate (v) at (0,-1);
        \draw [dotted, thick] (in) -- (x);
        \draw [dotted, thick] (x) -- (out);
        \draw [photon2] (x) -- (v) node [midway, left, black] {$q\!\downarrow$};
        \draw [photon] (v) -- (gin) node [left] {$p_1$} node [midway,above,sloped] {$\rightarrow$};
        \draw [photon] (v) -- (gout) node [right] {$p_2$} node [midway,above,sloped] {$\rightarrow$};
        \draw [gray,fill=white,thick] (x) circle (.2) node {\tiny$2$};
        \draw [fill] (v) circle (.08);
    \end{tikzpicture} = \ii\hat\delta(q\cdot v)\kappa^4M^2\int_\ell\frac{\hat\delta(\ell\cdot v)N_\text{iv}(\varepsilon_1,\varepsilon_2,\ell)}{\ell^2(q - \ell)^2}.
\end{equation}
Lastly, there is diagram (v) with two occurrences of the first-order vertex where the energy flows through a graviton line,
\begin{equation}
    \begin{tikzpicture}[baseline={(current bounding box.center)}]
        \coordinate (in) at (-1.25,0);
        \coordinate (out) at (1.25,0);
        \coordinate (x1) at (-.6,0);
        \coordinate (x2) at (.6,0);
        \coordinate (y1) at (-.6,-1);
        \coordinate (y2) at (.6,-1);
        \coordinate (g1) at (-1.25,-1.6);
        \coordinate (g2) at (1.25,-1.6);
        \coordinate (v) at (0,-.8);
        \draw [dotted, thick] (in) -- (out);
        \draw [photon2] (x1) -- (y1) node [midway, left, black] {$\ell\downarrow$};
        \draw [photon2] (x2) -- (y2) node [midway, right, black] {$\downarrow q-\ell$};
        \draw [photon] (y1) -- (y2) node [midway, below] {$\stackrel{\displaystyle\rightarrow}{p_1+\ell}$};
        \draw [photon] (y1) -- (g1) node [left] {$p_1$} node [midway,above,sloped] {$\rightarrow$};
        \draw [photon] (y2) -- (g2) node [right] {$p_2$} node [midway,above,sloped] {$\rightarrow$};
        \draw [gray,fill=white,thick] (x1) circle (.2) node {\tiny$1$};
        \draw [fill] (y1) circle (.08);
        \draw [gray,fill=white,thick] (x2) circle (.2) node {\tiny$1$};
        \draw [fill] (y2) circle (.08);
    \end{tikzpicture} = \ii\hat\delta(q\cdot v)\kappa^4M^2\int_\ell\frac{\hat\delta(\ell\cdot v)N_\text{v}(\varepsilon_1,\varepsilon_2,\ell)}{\ell^2(p_1+\ell)^2(q - \ell)^2}.
\end{equation}
All the diagrams have unit symmetry factor. Upon adding contributions from all diagrams, we find an integrand such that every integral that appears is a member of the integral family,
\begin{equation}
    K_{\nu_1,\nu_2,\nu_3,\lambda_1,\lambda_2} = \tilde\mu^{2\epsilon}\int_\ell\frac{\hat\delta(\ell\cdot v)(\varepsilon_1\cdot\ell)^{\lambda_1}(\varepsilon_2\cdot\ell)^{\lambda_2}}{[\ell^2]^{\nu_1}[(p_1 + \ell)^2]^{\nu_2}[(q - \ell)^2]^{\nu_3}}.
\end{equation}
Using integration-by-parts identities, implemented using \texttt{LiteRed} \cite{Lee:2012cn, Lee:2013mka}, we can express every integral in this family as a linear combination of the three master integrals \cite{Laporta:2000dsw},
\begin{equation}
    \mathcal{K}_1 = K_{0,1,0,0,0}, \qquad \mathcal{K}_2 = K_{1,0,1,0,0}, \qquad \mathcal{K}_3 = K_{1,1,1,0,0},
\end{equation}
and express the total second-order post-Minkowskian amplitude in this basis of integrals. As mentioned above, we also perform the computation in the weak-field expansion where the contributing diagrams and their respective symmetry factors are given in eq.\,\eqref{eq:flat2pm}. Naturally, the same integral family appears in this computation, and we find, reassuringly, complete agreement between the expansion coefficients calculated using curved space diagrams or flat space diagrams.

\subsection{Result}
We have computed the amplitude and verified its gauge-invariance in $d$ dimensions, which is ensured by highly non-trivial cancellations occurring between the diagrams. The $d$-dimensional result is complicated, so here we report only the expansion around $d = 4-2\epsilon$. In appendix \ref{app:mis}, we show the $\epsilon$-expansion of the master integrals to be
\begin{subequations}
\begin{align}
    \mathcal{K}_1 &= -\frac{\ii\omega}{4\pi}\Bigg[1 + \epsilon\bigg(\ii\pi + 2 - \log\frac{4\omega^2}{\mu^2}\bigg)\Bigg] + \mathcal{O}(\epsilon^2), \\
    \mathcal{K}_2 &= \frac{1}{8\abs{q}} + \mathcal{O}(\epsilon), \\
    \mathcal{K}_3 &= -\frac{\ii}{8\pi q^2\omega}\Bigg[\frac{1}{\epsilon} - \log\frac{-q^2}{\mu^2}\Bigg] + \mathcal{O}(\epsilon),
\end{align}
\end{subequations}
where we have included terms up to the order needed to determine the finite part of the amplitude. Expanded in the basis of master integrals, the amplitude takes the form,
\begin{equation}
    \ii\hat\delta(q\cdot v)\mathcal{M}^\text{2PM}(p_1,p_2) = \ii\hat\delta(q\cdot v)\kappa^4M^2\sum_{i=1}^3c_i\mathcal{K}_i,
\end{equation}
where the gauge-invariant master integral coefficients, $c_i$, are polynomials in dot products of the graviton momenta, polarizations, and the worldline velocity, with every term naturally restricted to be linear in each polarization tensor. Products of the gauge-invariant combinations,
\begin{subequations}
\begin{align}
    \mathsf{F}_1 &= \frac{v\cdot f_1\cdot f_2\cdot v}{\omega^2}, \\
    \mathsf{F}_2 &= \frac{(v\cdot f_1\cdot p_2)(v\cdot f_2\cdot p_1)}{\omega^4},
\end{align}
\end{subequations}
span this space, where $f_1$ and $f_2$ were defined in section \ref{sec:1pm}. We can therefore express the coefficients in the gauge-invariant basis, $\{\mathsf{F}_1^2, \mathsf{F}_2^2, \mathsf{F}_1\mathsf{F}_2\}$ \cite{Feng:2020jck}. The explicit expressions for the master integral coefficients can be found in table \ref{tab1}, where we use the ratio, $\kinpar = -q^2/4\omega^2$, defined in appendix \ref{app:mis}. In appendix \ref{app:IR}, we verify that the coefficients match those obtained from heavy-mass effective field theory.
\begin{table}
    \centering
    \begin{tabular}{cc} \toprule
       $c_1^{(-1)}$  & $\frac{1}{48\epsilon}\bigg[\frac{5\kinpar - 6}{2(y-1)^2}\mathsf{F}_1^2 - \frac{2 + \kinpar}{8(y-1)^4}\mathsf{F}_2^2 + \frac{\kinpar - 2}{(y-1)^3}\mathsf{F}_1\mathsf{F}_2\bigg]$  \\ \midrule
         $ c_1^{(0)} $ & $\frac{1}{288}\bigg[\frac{4\kinpar - 3}{(y-1)^2}\mathsf{F}_1^2 + \frac{4 + 17\kinpar}{8(y-1)^4}\mathsf{F}_2^2 + \frac{7 - 2\kinpar}{(y-1)^3}\mathsf{F}_1\mathsf{F}_2\bigg]$ \\ \midrule
        $ c_2^{(0)} $ & $\frac{\omega^2}{384}\bigg[\frac{45 + 15\kinpar - 45\kinpar^2 + \kinpar^3}{(y-1)^2}\mathsf{F}_1^2+ \frac{9+39\kinpar - \kinpar^2 + \kinpar^3}{4(y-1)^4}\mathsf{F}_2^2 + \frac{27 + 27\kinpar - 23\kinpar^2 + \kinpar^3}{(y-1)^3}\mathsf{F}_1\mathsf{F}_2\bigg]$\\ \midrule
        $c_3^{(0)}$ & $\frac{\omega^4}{6}\bigg[\frac{2\kinpar^2 - 3}{2(y-1)^2}\mathsf{F}_1^2 - \frac{2\kinpar + \kinpar^2}{8(y-1)^4}\mathsf{F}_2^2 + \frac{\kinpar^2 - 2\kinpar}{(y-1)^3}\mathsf{F}_1\mathsf{F}_2\bigg]$\\ \midrule
        $c_3^{(1)}$ & $\frac{\epsilon\omega^4}{36}\bigg[\frac{31\kinpar^2 - 36\kinpar}{(y-1)^2}\mathsf{F}_1^2 + \frac{3 - 8\kinpar - 10\kinpar^2}{8(y-1)^4}\mathsf{F}_2^2 + \frac{9 - 29\kinpar + 13\kinpar^2}{(y-1)^3}\mathsf{F}_1\mathsf{F}_2\bigg]$ \\ \bottomrule
    \end{tabular}
    \caption{The master integral coefficients for the second post-Minkowskian order Compton amplitude. We denote by $c_i^{(j)}$ the $\epsilon^j$ piece of $c_i$. 
    }
    \label{tab1}
\end{table}
In terms of the functions appearing in the master integrals, we obtain
\begin{equation}
    \mathcal{M}^\text{2PM}(p_1,p_2) = \frac{d_\text{IR}}{\epsilon} + d_1\log\frac{4\omega^2}{\mu^2} + d_2\log\frac{-q^2}{\mu^2} + d_\text{Im} \ii + \frac{d_q}{\abs{q}} + d_R,
\label{eq:result}\end{equation}
finding $-d_1 - d_2 = d_\text{IR}$ (see eq.\,\eqref{eq:d1d2}) so that the amplitude can equivalently be written as
\begin{equation}
    \mathcal{M}^\text{2PM}(p_1,p_2) = \frac{d_\text{IR}}{\epsilon} + d_\text{IR}\log\frac{\mu^2}{M^2} + d_1\log\frac{4\omega^2}{M^2} + d_2\log\frac{-q^2}{M^2} + d_\text{Im} \ii + \frac{d_q}{\abs{q}} + d_R.
\label{eq:result2}\end{equation}
We find
\begin{equation}
    d_\text{IR} = -\frac{\ii \kappa^2M\omega}{16\pi}\mathcal{M}^\text{1PM}(p_1,p_2),
\end{equation}
which provides an important check of the amplitude coming from Weinberg's soft graviton theorem~(see also \cite{Green:1983hw,DiVecchia:2025rdq} from string theory), where $d_\text{IR}$ is proportional to the first post-Minkowskian order amplitude~\cite{Weinberg:1965nx} (see also recent discussions for the waveform amplitude~\cite{Brandhuber:2023hhy,Herderschee:2023fxh,Georgoudis:2023lgf,Caron-Huot:2023vxl}). For $d_1$, $d_2$, and $d_\text{Im}$ we find
\begin{equation}
\begin{aligned}
    d_1 &= \frac{\ii\kappa^4M^2\omega}{3\times2^6\pi}\bigg[\frac{5\kinpar-6}{2(\kinpar - 1)^2}\mathsf{F}_1^2 - \frac{2+\kinpar}{8(\kinpar - 1)^4}\mathsf{F}_2^2 - \frac{2 - \kinpar}{(\kinpar - 1)^3}\mathsf{F}_1\mathsf{F}_2\bigg], \\
    d_2 &= \frac{\ii\kappa^4M^2\omega}{3\times2^6\pi}\bigg[\frac{3 - 2\kinpar^2}{2(\kinpar - 1)^2\kinpar}\mathsf{F}_1^2 + \frac{2+\kinpar}{8(\kinpar - 1)^4}\mathsf{F}_2^2 + \frac{2 - \kinpar}{(\kinpar - 1)^3}\mathsf{F}_1\mathsf{F}_2\bigg], \\
    d_\text{Im} &= \frac{\kappa^4M^2\omega}{3\times2^6\pi}\bigg[\frac{-1}{2(\kinpar - 1)}\mathsf{F}_1^2 - \frac{1 + 5\kinpar}{16(\kinpar - 1)^3\kinpar}\mathsf{F}_2^2 - \frac{3 - \kinpar}{2(\kinpar - 1)^2\kinpar}\mathsf{F}_1\mathsf{F}_2\bigg].
\end{aligned}
\label{eq:d1d2}\end{equation}
The terms with coefficients $d_1$, $d_2$, and $d_\text{IR}$ constitute the imaginary part of the amplitude, which is constrained entirely through the unitarity of the S-matrix (see appendix \ref{app:IR}). For the coefficient of $\abs{q}^{-1}$, we find
\begin{align}
    d_q &= \frac{\kappa^4M^2\omega^2}{3\times2^6}\bigg[\frac{45 + 15\kinpar - 45\kinpar^2 + \kinpar^3}{16(\kinpar - 1)^2}\mathsf{F}_1^2 \notag\\ &\hspace{2.5em}+ \frac{9 + 39\kinpar - \kinpar^2 + \kinpar^3}{64(\kinpar - 1)^4}\mathsf{F}_2^2 + \frac{27 + 27\kinpar - 23\kinpar^2 + \kinpar^3}{16(\kinpar - 1)^3}\mathsf{F}_1\mathsf{F}_2\bigg],
\end{align}
and, lastly, we determine the remaining rational part to be
\begin{equation}
    d_R = -\ii\pi d_1.
\end{equation}
We have already confirmed that the amplitude has the expected infrared behavior. As another check of the result, we can isolate the piece contributing to the classical massless scalar bending angle by extracting from eq.\,\eqref{eq:result} the coefficient of $(\varepsilon_1\cdot\varepsilon_2)^2$ (see e.g. \cite{Cheung:2017ems}). In the limit $\abs{q} \ll \omega \ll M$, only $d_q$ and $d_2$ contribute, giving
\begin{equation}
    2M\mathcal{M}^\text{2PM}\Big\vert_{(\varepsilon_1\cdot\varepsilon_2)^2} \simeq \frac{15\kappa^4M^3\omega^2}{512\abs{q}} - \frac{\ii\kappa^4M^3\omega^3}{16\pi q^2}\log\frac{-q^2}{M^2}.
\label{eq:geom-optics}\end{equation}
Here, we have multiplied by $2M$ to enable a comparison with other results in the literature (see the discussion on normalization in section \ref{sec:1pm}). We note that the presence of the imaginary phase term in eq.\,\eqref{eq:geom-optics} is necessary to obtain the consistent infrared behavior seen in our computation. It does not affect the result for classical physics, such as the bending angle, as demonstrated in \cite{Bjerrum-Bohr:2014zsa}.

\section{Conclusion}
\label{sec:conc}
In this work, we have employed Feynman rules derived in curved spacetime to calculate the second post-Minkowskian order correction to the gravitational Compton amplitude and conducted a series of non-trivial checks to validate the accuracy of our results, including comparison to equivalent computations in flat space. Our findings demonstrate that the application of curved space Feynman rules allows for a reorganization and, to some extent, a simplification of traditional Feynman rule calculations by leveraging exact-in-$G$ classical information from general relativity. In particular, we have elaborated on how expansions around a Schwarzschild-Tangherlini background enable the resummation of an infinite series of potential graviton diagrams, as illustrated in eq.\,\eqref{eq:se-wfe-correspondence}, and as previously discussed in \cite{Kosmopoulos:2023bwc, Cheung:2023lnj,Cheung:2024byb}. Notably, the dimensional regularization approach eliminates diagrams that involve self-energy from the overall diagrammatic expansion.

While these advantages may seem modest and the prospects for post-Minkowskian resummation appear tentative, particularly given that the two-point graviton vertex in the curved space expansion (as shown in eq.\,\eqref{eq:2-pt-expanded-diagram}) corresponds to an infinite series of loop diagrams, they are not completely out of reach as demonstrated by the resummation of the graviton one-point function sourced by a massive scalar \cite{PhysRevLett.132.251603, Mougiakakos:2024lif}. The methods employed in these papers, other novel approaches to resummation, and even the possibilities put forth by alternative coordinate choices, such as the Kerr-Schild coordinate gauge \cite{Kerr:1965wfc} for the background metric, could potentially facilitate progress in this area. The Kerr-Schild gauge could be especially relevant when considering higher-order corrections, as the metric in this gauge, being linear in $G$, would allow the curved $n$-point graviton vertex to truncate at a finite order in $G$.  Currently, black hole perturbation theory~\cite{Regge:1957td,Zerilli:1970se,Teukolsky:1972my,Teukolsky:1973ha} is instrumental in understanding both the quasinormal ringdown~\cite{Vishveshwara:1970zz,Chandrasekhar:1975zza,Leaver:1985ax,Konoplya:2003ii,Berti:2009kk}, finite size effects, and the self-force expansion of the inspiral dynamics~\cite{Mino:1996nk,Quinn:1996am,Barack:2009ux,Hinderer:2008dm} of compact binary mergers. Methods for Compton amplitudes in curved spacetimes may be useful in studying these phenomena in the language of quantum field theory \cite{Ivanov:2024sds}.

Natural future research directions involve computing and analyzing the classical Compton amplitude at higher post-Minkowskian orders. Investigating their infrared behavior and the master integrals that arise at these higher post-Minkowskian orders represents a critical and intriguing avenue for further exploration of the amplitude structures relevant to classical gravitational physics.

\acknowledgments{We thank Paolo Di Vecchia, Mikhail Solon, and Pierre Vanhove for discussions and comments on the draft. The work of N.E.J.B.-B., G.C., N.S. was supported in part by DFF grant 1026-00077B, The Center of Gravity, which is a Center of Excellence funded by the Danish National Research Foundation under grant No. 184 and in part by VILLUM Foundation (grant no. VIL37766). G.C. has also received funding in part from the European Union Horizon 2020 research and innovation program under the Marie Sklodowska-Curie grant agreement No. 847523 INTERACTIONS.}

\appendix
\section{Fourier transform}\label{app:fourier}
We review the Fourier transform needed to compute the two-point vertex from eq.\,\eqref{eq:2-point-se} in the curved space expansion. The Fourier transform contains an integration along the velocity of the black hole, $v^\mu$, which trivially evaluates to a delta function,
\begin{align}
    \int\mathrm{d}^dx\,\ee^{\ii q\cdot x}\frac{1}{\abs{x_\perp}^n} &= \int\mathrm{d}^{d-1}x_\perp\mathrm{d}x_\parallel\,\ee^{\ii q\cdot x}\frac{1}{\abs{x_\perp}^n} \notag\\
    &= \hat\delta(q\cdot v)\int\mathrm{d}^{d-1}x_\perp\,\ee^{\ii q\cdot x_\perp}\frac{1}{\abs{x_\perp}^n}.
\end{align}
Then, using $\abs{x_\perp}^n = (\abs{x_\perp}^2)^\frac{n}{2}$ and introducing a Schwinger parameter, we get
\begin{align}
    \int\mathrm{d}^{d-1}x_\perp\,\ee^{\ii q\cdot x_\perp}\frac{1}{\abs{x_\perp}^n} &= \frac{1}{\Gamma(\frac{n}{2})}\int_0^\infty\mathrm{d}\alpha\,\alpha^{\frac{n}{2}-1}\int\mathrm{d}^{d-1}x_\perp\,\ee^{-\alpha\abs{x_\perp}^2+\ii q\cdot x_\perp} \notag\\
    &= \frac{\pi^\frac{d-1}{2}}{\Gamma(\frac{n}{2})}\int_0^\infty\mathrm{d}\alpha\,\alpha^{\frac{n-d-1}{2}}\exp\frac{q^2}{4\alpha} \notag\\
    &= \frac{(4\pi)^\frac{d-1}{2}\Gamma(\frac{d-1-n}{2})}{2^n\Gamma(\frac{n}{2})}\frac{1}{(-q^2)^\frac{d-1-n}{2}}.
\end{align}
We can recover the result used in the main text by substituting $n = d - 3$. To wit,
\begin{align}
    \int\mathrm{d}^dx\,\ee^{\ii q\cdot x}\frac{1}{\abs{x_\perp}^{d-3}} &= \frac{4\pi^\frac{d-1}{2}}{\Gamma(\frac{d-3}{2})}\frac{\hat\delta(q\cdot v)}{-q^2} \notag\\
    &= \hat\delta(q\cdot v)\frac{(d-3)\Omega_{d-2}}{-q^2},
\end{align}
where we used $\Gamma(n+1) = n\Gamma(n)$ and $\Omega_{d-2}$ was defined in eq.\,\eqref{eq:metric-scales}.

For terms at order $\kappa^{2n}$, we need
\begin{equation}
    \int\mathrm{d}^dx\,\ee^{\ii q\cdot x}\frac{\Lambda_d}{\abs{x_\perp}^{d-3}} = \hat\delta(q\cdot v)\frac{\Lambda_d(d-3)\Omega_{d-2}}{-q^2}.
\label{eq:fouriertrans}\end{equation}
Notice that the prefactor combines nicely to give
\begin{equation}
    \Lambda_d(d-3)\Omega_{d-2} = \frac{\kappa^2M}{2}\frac{d-3}{d-2}.
\end{equation}
Using the fact that a product in position space is a convolution in momentum space,
\begin{equation}
    \int\mathrm{d}^dx\,\ee^{\ii q\cdot x}\frac{\Lambda_d^n}{\abs{x_\perp}^{n(d-3)}} = \hat\delta(q\cdot v)\frac{(\kappa^2M)^n}{2^n}\frac{(d-3)^n}{(d-2)^n}\int_{\ell_1,\ldots,\ell_{n-1}}\frac{(-1)^n\hat\delta(\ell_1\cdot v)\cdots\hat\delta(\ell_{n-1}\cdot v)}{\ell_1^2\cdots\ell_{n-1}^2(q-\ell_{1\cdots (n-1)})^2},
\label{eq:nth-order-cut triangle-integral}\end{equation}
where $\ell_{1\cdots (n-1)} = \sum_{i=1}^{n-1}\ell_i$. These types of integrals can be evaluated exactly for any $n$.

\section{Master integrals}\label{app:mis}
In the second post-Minkowskian order computation, we encounter the one-loop integral family,
\begin{equation}
    K_{\nu_1,\nu_2,\nu_3,\lambda_1,\lambda_2} = \tilde\mu^{2\epsilon}\int\frac{\mathrm{d}^d\ell}{(2\pi)^d}\frac{\hat\delta(\ell\cdot v)(\varepsilon_1\cdot\ell)^{\lambda_1}(\varepsilon_2\cdot\ell)^{\lambda_2}}{[\ell^2]^{\nu_1}[(p_1 + \ell)^2 + \ii0]^{\nu_2}[(q - \ell)^2]^{\nu_3}},
\end{equation}
where we omit the $\ii0$ prescription on the first and last propagators, as the delta function forces these momenta to be potential modes which are never on-shell. We find that three master integrals span this family,
\begin{equation}
    \mathcal{K}_1 = K_{0,1,0,0,0}, \qquad \mathcal{K}_2 = K_{1,0,1,0,0}, \qquad \mathcal{K}_3 = K_{1,1,1,0,0},
\end{equation}
the cut bubble, cut triangle, and cut box, respectively (see figure \ref{fig:mis}). The first two are simple to evaluate. After integrating out the delta function, the cut bubble is
\begin{equation}
    \mathcal{K}_1 = \tilde\mu^{2\epsilon}\int\frac{\mathrm{d}^{d-1}\ell_\perp}{(2\pi)^{d-1}}\frac{1}{(\ell_\perp + p_{1\perp})^2 + \omega^2 + \ii0}.
\end{equation}
Shifting the loop momentum and using the primitive,
\begin{equation}
    \int\frac{\mathrm{d}^nk}{(2\pi)^n}\frac{1}{k^2 + \Delta} = \frac{\Gamma(1-\frac{n}{2})}{(4\pi)^\frac{n}{2}}\Delta^{\frac{n}{2}-1},
\end{equation}
continued to $\Delta < 0$ and with $n = d-1$, we obtain
\begin{align}
    \mathcal{K}_1 &= -\frac{\Gamma(\epsilon-\frac12)}{(4\pi)^{\frac32-\epsilon}}\tilde\mu^{2\epsilon}(-\omega^2 - \ii0)^{\frac12-\epsilon} \\
    &= -\frac{\ii\omega}{4\pi}\Bigg[1 + \epsilon\bigg(\ii\pi + 2 - \log\frac{4\omega^2}{\mu^2}\bigg)\Bigg] + \mathcal{O}(\epsilon).
\end{align}
The cut triangle is, upon integrating out the delta function, a standard integral,
\begin{align}
    \mathcal{K}_2 &= \tilde\mu^{2\epsilon}\int\frac{\mathrm{d}^{d-1}\ell_\perp}{(2\pi)^{d-1}}\frac{1}{\ell_\perp^2(q_\perp - \ell_\perp)^2} \notag\\
    &= \frac{\Gamma(\frac12 - \epsilon)^2\Gamma(\frac12 + \epsilon)}{(4\pi)^{\frac32 - \epsilon}\Gamma(1 - 2\epsilon)}\frac{\tilde\mu^{2\epsilon}}{(-q^2)^{\frac12+\epsilon}} \notag\\
    &= \frac{1}{8\abs{q}} + \mathcal{O}(\epsilon),
\end{align}
where we used $q_\perp = q$. Finally, we come to the cut box, which, after integrating out the delta function, is
\begin{figure}[t]
    \centering
    \begin{tikzpicture}[baseline={(current bounding box.center)}]
        \coordinate (v1) at (-1.3,0);
        \coordinate (v2) at (1.3,0);
        \coordinate (v3) at (-1.3,-1.5);
        \coordinate (v4) at (1.3,-1.5);
        \coordinate (v5) at (0,.1);
        \coordinate (v6) at (0,-.5);
        \draw [zUndirected,looseness=2] (v1) to[in=180+60,out=-60] (v2);
        \draw [zUndirected,looseness=2] (v3) to[in=180-60,out=60] (v4);
        \draw [zUndirected,red] (v5) -- (v6);
    \end{tikzpicture}\hspace{1cm}
    \begin{tikzpicture}[baseline={(current bounding box.center)}]
        \coordinate (v1) at (-1.3,0);
        \coordinate (v2) at (1.3,0);
        \coordinate (v3) at (-.6,-2);
        \coordinate (v4) at (.6,-2);
        \coordinate (v5) at (-.9,-.41);
        \coordinate (v6) at (.9,-.41);
        \coordinate (v7) at (0,-.1);
        \coordinate (v8) at (0,-.7);
        \draw [zUndirected] (v1) -- (v4);
        \draw[zUndirected] (v2) -- (v3);
        \draw[zUndirected] (v5) -- (v6);
        \draw[zUndirected,red] (v7) -- (v8);
    \end{tikzpicture}\hspace{1cm}
    \begin{tikzpicture}[baseline={(current bounding box.center)}]
        \coordinate (out1) at (-1.3,0);
        \coordinate (out2) at (1.3,0);
        \coordinate (out3) at (-1.3,-2);
        \coordinate (out4) at (1.3,-2);
        \coordinate (in1) at (-.8,-.5);
        \coordinate (in2) at (.8,-.5);
        \coordinate (in3) at (-.8,-2+.5);
        \coordinate (in4) at (.8,-2+.5);
        \coordinate (v1) at (0,-.2);
        \coordinate (v2) at (0,-.8);
        \draw[zUndirected] (out1) -- (in1);
        \draw[zUndirected] (out2) -- (in2);
        \draw[zUndirected] (out3) -- (in3);
        \draw[zUndirected] (out4) -- (in4);
        \draw[zUndirected] (in1) -- (in2);
        \draw[zUndirected] (in2) -- (in4);
        \draw[zUndirected] (in3) -- (in4);
        \draw[zUndirected] (in3) -- (in1);
        \draw[zUndirected,red] (v1) -- (v2);
    \end{tikzpicture}
    \caption{The cut bubble $\mathcal{K}_1$, the cut triangle $\mathcal{K}_2$, and the cut box $\mathcal{K}_3$. The red line signifies the propagator that is cut.}
    \label{fig:mis}
\end{figure}
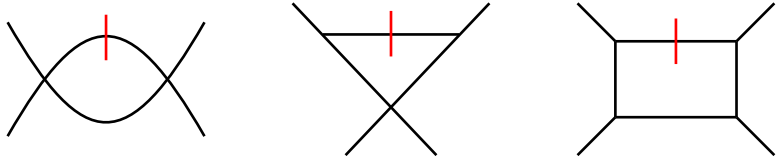
\begin{equation}
    \mathcal{K}_3 = \tilde\mu^{2\epsilon}\int\frac{\mathrm{d}^{d-1}\ell_\perp}{(2\pi)^{d-1}}\frac{1}{\ell_\perp^2[(p_{1\perp} + \ell_\perp)^2 + \omega^2 + \ii0](q_\perp - \ell_\perp)^2}.
\end{equation}
We find it convenient to evaluate it with the method of differential equations \cite{Kotikov:1990kg,Gehrmann:1999as,Smirnov:2008iw,larsen2016integration,Lee:2014ioa}. Naïvely, the integral can depend independently on the energy, $\omega$ and, by Lorentz invariance, the squared momentum transfer, $q_\perp^2$. However, by extracting the mass dimension from all the momenta,
\begin{equation}
    P \to \frac{P}{\omega}, \quad\text{where}\quad P = \ell_\perp, p_{1\perp}, q_\perp,
\label{eq:mass-rescaling}\end{equation}
$\omega$ appears only as a prefactor, implying that the integral depends only on the ratio, $\kinpar$, of the two kinematic invariants,
\begin{equation}
    \mathcal{K}_3(\omega,q^2) = \omega^{-3-2\epsilon}\tilde\mu^{2\epsilon}\mathsf{K}_3(\kinpar), \qquad \kinpar = \frac{-q^2}{4\omega^2}.
\end{equation}
We can see from
\begin{equation}
    q_\perp^2 = p_{1\perp}^2 + p_{2\perp}^2 - 2p_{1\perp}\cdot p_{2\perp} = -2\omega^2(1 - \cos\theta),
\end{equation}
where $\theta$ is the angle between $p_{1\perp}$ and $p_{2\perp}$, that the physical region is $\kinpar \in [0,1]$, with $\kinpar \to 0$ being the \emph{forward-scattering limit}. The differential equation is derived using \texttt{LiteRed},
\begin{equation}
    \frac{\mathrm{d}}{\mathrm{d}\kinpar}\mathsf{K}_3(\kinpar) = \frac{1}{\kinpar(\kinpar - 1)}\bigg(\frac{1-2\epsilon}{8}\mathsf{K}_1(\kinpar) - \frac{\epsilon}{2}\mathsf{K}_2(\kinpar) - (\kinpar - 1 - \epsilon)\mathsf{K}_3(\kinpar)\bigg),
\end{equation}
where $\mathsf{K}_1$ and $\mathsf{K}_2$ are the cut bubble and cut triangle with their $\omega$ dependence properly extracted using eq.\,\eqref{eq:mass-rescaling}. This differential equation has a solution in terms of hypergeometric functions,
\begin{equation}
\begin{aligned}
    \mathsf{K}_3(\kinpar) &= (1 - \kinpar)^\epsilon \\
    &\times\Bigg(\kinpar^{-1-\epsilon}C(\epsilon) + A_{\mathsf{K}_1}(\epsilon)\,{}_2F_1\Big[\begin{gathered}\scriptstyle 1{+}\epsilon,\,1{+}\epsilon, \\[-.5em]\scriptstyle2+\epsilon\end{gathered}; \kinpar\Big] + \kinpar^{-\frac12-\epsilon}A_{\mathsf{K}_2}(\epsilon)\,{}_2F_1\Big[\begin{gathered}\scriptstyle \frac12,\,1{+}\epsilon, \\[-.5em] \scriptstyle\frac32\end{gathered}; \kinpar\Big]\Bigg),
\end{aligned}
\end{equation}
where $C(\epsilon)$ is an integration constant, and
\begin{equation}
    A_{\mathsf{K}_1}(\epsilon) = \ii(-1)^{-\epsilon}\frac{\Gamma(\epsilon-\frac12)}{(4\pi)^{\frac32-\epsilon}}, \qquad A_{\mathsf{K}_2}(\epsilon) = \frac{1}{2^{1+2\epsilon}}\frac{\Gamma(\frac12 - \epsilon)^2\Gamma(\frac12 + \epsilon)}{(4\pi)^{\frac32 - \epsilon}\Gamma(1 - 2\epsilon)}.
\end{equation}
To determine the boundary constant, notice that the asymptotic expansion of the integral in the forward-scattering limit is
\begin{equation}
    \mathsf{K}_3(\kinpar) \sim \frac{C(\epsilon)}{\kinpar^{1+\epsilon}} \quad\text{as}\quad \kinpar\to0.
\label{eq:mi3-asym}\end{equation}
We can determine this asymptotic expansion with the method of regions. We follow the approach proposed in \cite{Pak:2010pt,Semenova:2018cwy} (see also appendix A in \cite{Brandhuber:2021eyq}), which allows us to perform the expansion at the level of Feynman parameters. Using Feynman parameters,
\begin{align}
    \mathsf{K}_3(\kinpar) &= -\frac{\Gamma(\frac32+\epsilon)}{(4\pi)^{\frac32-\epsilon}}\int_{\alpha_i\geq0}\mathrm{d}^3\alpha\,\delta\Big(1-\sum_{i\in S}\alpha_i\Big)\frac{(\alpha_1 + \alpha_2 + \alpha_3)^{2\epsilon}}{(4\kinpar\alpha_1\alpha_3-\alpha_2^2 - \ii0)^{\frac32+\epsilon}} \notag\\
    &= -\frac{\Gamma(\frac32+\epsilon)}{(4\pi)^{\frac32-\epsilon}}\int_{\alpha_i\geq0}\mathrm{d}^2\alpha\,\frac{(1 + \alpha_1 + \alpha_3)^{2\epsilon}}{(4\kinpar\alpha_1\alpha_3-1 - \ii0)^{\frac32+\epsilon}},
\end{align}
where we used the Cheng-Wu theorem to choose $S = \{2\}$ \cite{Cheng:1969eh}. The method instructs us to find all variable substitutions, $\alpha_i\to\kinpar^{v_i}\alpha_i$, that result in a non-scaleless integral with $\kinpar$-homogeneous Symanzik polynomials after expanding the integrand in $\kinpar$. We find that the only such substitution is $\alpha_i \to \kinpar^{-\frac12}\alpha_i$, which leads to an integral we can readily evaluate,
\begin{equation}
\begin{array}{cc}
    \begin{aligned}
        \mathsf{K}_3(\kinpar) &\sim -\frac{1}{\kinpar^{1+\epsilon}}\frac{\Gamma(\frac32+\epsilon)}{(4\pi)^{\frac32-\epsilon}}\int_{\alpha_i\geq0}\mathrm{d}^2\alpha\,\frac{(\alpha_1 + \alpha_3)^{2\epsilon}}{(4\alpha_1\alpha_3-1-\ii0)^{\frac32+\epsilon}} \\
        &= (-1-\ii0)^{\frac{5}{2}-\epsilon}\frac{\pi}{4}\frac{\big(\ii-\cot\pi\epsilon\big)\Gamma\big(\frac{1}{2}+\epsilon\big)\cos\pi\epsilon}{(4\pi)^{\frac{3}{2}-\epsilon}}\frac{1}{y^{1+\epsilon}},
    \end{aligned} & \qquad\text{as}\quad \kinpar\to0.
\end{array}
\end{equation}
Comparing this with eq.\,\eqref{eq:mi3-asym}, we can identify $C(\epsilon)$. Finally, we may determine the $\epsilon$-expansion of the cut box to be
\begin{equation}
   \mathcal{K}_3 = -\frac{\ii}{8\pi q^2\omega}\Bigg[\frac{1}{\epsilon} - \log\frac{-q^2}{\mu^2}\Bigg] + \mathcal{O}(\epsilon).
\end{equation}
\section{Scaleless contributions to the worldline action}\label{app:scaleless}
To prove the scaleless-ness of non-trivial contributions from the background metric evaluated at the position of the massive body, we start by manipulating $S_\text{wl}[\bar\gamma,x]$ analogously to $S_\text{wl}[h,x]$ in section \ref{sec:Weak-field expansion}. We first Fourier transform the curved part of the metric,
\begin{equation}
    \bar\gamma_{\mu\nu}(x(t)) = \int_{\ell_1}\ee^{\ii\ell_1\cdot x(\tau)}\bar\gamma_{\mu\nu}(-\ell_1).
\label{eq:curved-piece-fourier}\end{equation}
Expanding the exponential in eq.\,\eqref{eq:curved-piece-fourier} and inserting yields an expression identical to what we obtained before, i.e.,
\begin{equation}
\begin{aligned}
    S_\text{wl}[\bar\gamma,x] &= - M\sum_{j=0}^\infty\frac{\ii^j}{j!}
    \int_{\ell_1,\omega_1,\ldots,\omega_j}\hat\delta(\ell_1\cdot v + \omega_{1\cdots j})
    \bar\gamma_{\mu\nu}(-\ell_1)\prod_{l=1}^jz^{\rho_l}(-\omega_l) \\
    &\times\Bigg[\frac12\prod_{n=1}^jk_{\rho_n}v^\mu v^\nu+
    \sum_{n=1}^j\prod_{m\neq n}^j\omega_nk_{\rho_m}\!
    v^{(\mu}\delta^{\nu)}_{\rho_n}+
    \sum_{n<m}^j\prod_{s\neq n,m}^j\omega_n\omega_mk_{\rho_s}
    \delta^{(\mu}_{\rho_n}\delta^{\nu)}_{\rho_m}\Bigg].
\end{aligned}
\label{eq:scaleless}\end{equation}
For the $z^j$ order piece of this equation, we can say that
\begin{equation}
    S_\text{wl}[\bar\gamma,x]\big\vert_{z^j} \propto \int_{\ell_1,\omega_1,\ldots,\omega_j}\prod_{l=1}^jz^{\rho_l}(-\omega_l)N^{\mu\nu}_{\rho_1\cdots\rho_j}[\ell_1,\{\omega_i\}_{i=1..j}]\hat\delta(\ell_1\cdot v + \omega_{1\cdots j})\gamma_{\mu\nu}(-\ell_1),
\end{equation}
where we collected the numerator factors in $N^{\mu\nu}_{\rho_1\cdots\rho_j}[\ell_1,\{\omega_i\}_{i=1..j}]$. As we showed above, $\bar\gamma_{\mu\nu}(-\ell_1)$ is given at order $\kappa^{2n}$ by the cut triangle-type $(n-1)$-loop integral in eq.\,\eqref{eq:nth-order-cut triangle-integral}. Thus, extracting the $\kappa^{2n}$-order piece of the above, we have
\begin{equation}
\begin{aligned}
    S_\text{wl}[\bar\gamma,x]\big\vert_{z^j\kappa^{2n}} \propto& \int_{\ell_1,\ldots,\ell_n,\omega_1,\ldots,\omega_j}\prod_{l=1}^jz^{\rho_l}(-\omega_l)N^{\mu\nu}_{\rho_1\cdots\rho_j}[\ell_1,\{\omega_i\}_{i=1..j}] \\
    &\hspace{1em}\times\frac{\hat\delta(\ell_1\cdot v + \omega_{1\cdots j})\hat\delta(\ell_1\cdot v)\hat\delta(\ell_2\cdot v)\cdots\hat\delta(\ell_{n}\cdot v)}{\ell_2^2\cdots\ell_n^2(\ell_1+\ell_{2\cdots n})^2}.
\end{aligned}
\end{equation}
Using $\hat\delta(\ell_1\cdot v)$, and substituting $\ell_1 \to \ell_1 - \ell_{2\cdots n}$, we are left with
\begin{equation}
\begin{aligned}
    S_\text{wl}[\bar\gamma,x]\big\vert_{z^j\kappa^{2n}} &\propto \int_{\omega_1,\ldots,\omega_j}\prod_{l=1}^jz^{\rho_l}(-\omega_l)\hat\delta(\omega_{1\cdots j}) \\
    &\times\int_{\ell_1,\ldots,\ell_n}\frac{\hat\delta(\ell_1\cdot v)\hat\delta(\ell_2\cdot v)\cdots\hat\delta(\ell_{n}\cdot v)N^{\mu\nu}_{\rho_1\cdots\rho_j}[\ell_1 - \ell_{2\cdots n},\{\omega_i\}_{i=1..j}]}{\ell_1^2\ell_2^2\cdots\ell_n^2}.
\end{aligned}
\end{equation}
When expanded, each piece of the above factorizes into manifestly scaleless integrals. To see this, we notice by inspecting eq.\,\eqref{eq:scaleless} that each term in the numerator, $N^{\mu\nu}_{\rho_1\cdots\rho_j}$, from the above expression will be proportional to products of the form,
\begin{equation}
    \prod_{i=1}^\iota(\ell_1 - \ell_{2\cdots n})_{\rho_{s_i}},
\label{eq:generic-numerator}\end{equation}
where $\iota = j-2, j-1, j$ corresponds to terms coming from the first, second, and third terms of eq.\,\eqref{eq:scaleless}, and $s_i = 1,\ldots,j$. When expanded, a term in eq.\,\eqref{eq:generic-numerator} will, in general contain $1\leq R\leq\iota$ different loop momenta, meaning that it can be written as
\begin{equation}
    \prod_{i=1}^R\prod_{s_i}\ell_{r_i\rho_{s_i}},
\end{equation}
where $\rho_{s_i}$ denotes the indices that $\ell_{r_i}$ has in the term. With this knowledge, we can deduce that any term in $S_\text{wl}[\bar\gamma,x]\big\vert_{z^j\kappa^{2n}}$ will be proportional to a product of integrals of the form
\begin{equation}
    \Bigg[\int_{\ell}\frac{\hat\delta(\ell\cdot v)}{\ell^2}\Bigg]^{n-R}\prod_{i=1}^R\int_{\ell_{r_i}}\frac{\hat\delta(\ell_{r_i}\cdot v)\prod_{s_i}\ell_{r_i\rho_{s_i}}}{\ell_{r_i}^2}
\end{equation}
These integrals are all scaleless, so $S_\text{wl}[\bar\gamma,x]$ vanishes in dimensional regularization. Diagrammatically, we can understand this fact as the vanishing of all diagrams where the worldline emits and absorbs a thick background ``graviton", e.g.,
\begin{equation}
    \begin{tikzpicture}[baseline={(current bounding box.center)}]
        \coordinate (in) at (-1.5,0);
        \coordinate (out) at (1.5,0);
        \coordinate (x1) at (-.8,0);
        \coordinate (x2) at (.8,0);
        \draw [dotted, thick] (in) -- (out);
        \draw [white] (x1) to [bend left=80] (x2);
        \draw [photon2] (x1) to [bend right=80] (x2);
        \node [gray,draw,thick,fill=white,circle,cross,minimum width=.3cm] at (x1) {};
        \node [gray,draw,thick,fill=white,circle,cross,minimum width=.3cm] at (x2) {};
    \end{tikzpicture}\,, \quad \begin{tikzpicture}[baseline={(current bounding box.center)}]
        \coordinate (in) at (-1.5,0);
        \coordinate (out) at (1.5,0);
        \coordinate (x1) at (-.8,0);
        \coordinate (x2) at (.8,0);
        \draw [dotted, thick] (in) -- (x1);
        \draw [dotted, thick] (x2) -- (out);
        \draw [zUndirected] (x1) -- (x2);
        \draw [white] (x1) to [bend left=80] (x2);
        \draw [photon2] (x1) to [bend right=80] (x2);
        \node [gray,draw,thick,fill=white,circle,cross,minimum width=.3cm] at (x1) {};
        \node [gray,draw,thick,fill=white,circle,cross,minimum width=.3cm] at (x2) {};
    \end{tikzpicture}\,, \quad \begin{tikzpicture}[baseline={(current bounding box.center)}]
        \coordinate (in) at (-1.5,0);
        \coordinate (out) at (1.5,0);
        \coordinate (x1) at (-.8,0);
        \coordinate (x2) at (.8,0);
        \draw [dotted, thick] (in) -- (x1);
        \draw [dotted, thick] (x2) -- (out);
        \draw [zUndirected] (x1) -- (x2);
        \draw [zUndirected] (x1) to [bend left=80] (x2);
        \draw [photon2] (x1) to [bend right=80] (x2);
        \node [gray,draw,thick,fill=white,circle,cross,minimum width=.3cm] at (x1) {};
        \node [gray,draw,thick,fill=white,circle,cross,minimum width=.3cm] at (x2) {};
    \end{tikzpicture}\,, \quad \text{etc.}
\end{equation}
The above proof for the vanishing of this term generalizes easily through the Fourier transform in eq.\,\eqref{eq:fouriertrans} to any metric, $g'$, which admits an expansion of the form
\begin{equation*}
    g'_{\mu\nu}(x) = \eta_{\mu\nu} + \sum_n\frac{g_{\mu\nu}'^{(n)}}{\abs{x_\perp}^{(d-3)n}}.
\end{equation*}
\section{Second post-Minkowskian order Compton amplitude from heavy-mass effective field theory}\label{app:IR}
In this section, we compare the second post-Minkowskian order Compton amplitude obtained from Feynman rules with the one from heavy-mass effective field theory in \cite{Brandhuber:2021kpo,Brandhuber:2021eyq}. 
The second post-Minkowskian order classical Compton amplitude can be constructed from the union of the following diagrams,
\begin{align}
\label{eq:zigRad}
  \begin{tikzpicture}[baseline={([yshift=-0.4ex]current bounding box.center)}]\tikzstyle{every node}=[font=\small]	
\begin{feynman}
    	 \vertex (p1) {};
    	 \vertex [right=1.2cm of p1] (b1) []{ $\mathbf{1}$};
    	  \vertex [right=1.5cm of b1] (b2) []{ $\mathbf{2}$};
    	 \vertex [right=1.2cm of b2] (p4){};
    	 \vertex [above=2.0cm of p1](p2){$v$};
    	 \vertex [right=1.2cm of p2] (u1) [HV]{H};
    	 \vertex [right=1.5cm of u1] (u2) [HV]{H};
    	  \vertex [right=1.2cm of u2](p3){};
    	   \vertex [right=0.75cm of u1] (cut1);
    	  \vertex [above=0.2cm of cut1] (cut1u);
    	  \vertex [below=0.1cm of cut1] (cut1b);
          \vertex [below=0.7cm of cut1b] (cut1bb);
    	  \diagram* {(u1)-- [photon,ultra thick,out=-45,in=-135,looseness=0.5,min distance=0.7cm,momentum'=\(\ell\)] (u2),
    	  (u1)--[photon,ultra thick,momentum'=\(p_1\)](b1), (u2)-- [photon,ultra thick,momentum=\(p_2\)] (b2), (p2) -- [fermion,thick] (u1)-- [fermion,thick] (u2)-- [fermion,thick] (p3), (cut1u)--[ red,thick] (cut1b), (cut1b)--[dashed,red,thick] (cut1bb)    	  };
    \end{feynman}  
    \end{tikzpicture} &&\cup &&
    \begin{tikzpicture}[baseline={([yshift=-0.4ex]current bounding box.center)}]\tikzstyle{every node}=[font=\small]	
\begin{feynman}
    	 \vertex (p1) {};
    	 \vertex [right=1.2cm of p1] (b1) []{ $\mathbf{1}$};
    	  \vertex [right=1.5cm of b1] (b2) []{ $\mathbf{2}$};
    	 \vertex [right=1.2cm of b2] (p4){};
    	 \vertex [above=2.0cm of p1](p2){$v$};
    	 \vertex [right=1.2cm of p2] (u1) [HV]{H};
    	 \vertex [right=1.5cm of u1] (u2) [HV]{H};
    	  \vertex [right=1.2cm of u2](p3){};
          \vertex [above=1.0cm of b1] (m1)[]{};
          \vertex [right=0.75cm of m1] (m2)[HV]{G};
           \vertex [right=0.75cm of m2] (m3)[]{};
    	   \vertex [right=0.75cm of u1] (cut1);
    	  \vertex [above=0.2cm of cut1] (cut1u);
    	  \vertex [below=0.2cm of cut1] (cut1b);
          \vertex [below=1cm of cut1b] (cut1bb);
    	  \diagram* {
    	  (u1)--[photon,ultra thick,momentum'=\(\ell+p_1\)](m2)--[photon,ultra thick](b1), (b2)--[photon,ultra thick](m2)--[photon,ultra thick,momentum'=\(\ell-p_2\)](u2), (p2) -- [fermion,thick] (u1)-- [fermion,thick] (u2)-- [fermion,thick] (p3), (cut1u)--[ red,thick] (cut1b), (cut1b)--[dashed,red,thick] (m1) , (cut1b)--[dashed,red,thick] (m3)     	  };
    \end{feynman}  
    \end{tikzpicture} . \nn\\
    (a)~~~~~~~~~~~~~~~&&~ &&(b)~~~~~~~~~~~~~~
\end{align}
where we follow the notation of \cite{Brandhuber:2021eyq}.\footnote{In particular, in contrast to the main text, both graviton momenta, $p_1$ and $p_2$, are taken outgoing.} The first diagram constructs all the terms with master integrals, $\mathcal{K}_1$ and $\mathcal{K}_3$, while the second diagram provides the terms with $\mathcal{K}_2$ and $\mathcal{K}_3$. The union of the two graphs means keeping only one copy of the overlapping $\mathcal{K}_3$ term and adding it together with the $\mathcal{K}_1$ and $\mathcal{K}_2$ terms.    After summation over the intermediate graviton state in the first graph, we get the integrand,
\begin{align}
	A_{\rm comp}^{(1,a)}(p_1,p_2,v)&=\Big(\int {d^D\ell\over (2\pi)^D } \delta(2Mv\mdot (\ell+p_1)) {1\over \ell^2}\sum_{h_\ell} A_{\rm comp}(p_1,\ell,v) A_{\rm comp}(-\ell,p_2,v)\nn\\
	&=\int {d^D\ell\over (2\pi)^D } \delta(2Mv\mdot (\ell+p_1)) {1\over \ell^2}\Bigg(-\frac{(D-3) M^4 \left(\ell\mdot f_1\mdot v\right)^2 \left(\ell\mdot f_2\mdot v\right)^2}{4 (D-2) \ell\mdot p_1 \ell\mdot p_2 \left(p_1\mdot v\right)^4}\nn\\
	&+\frac{M^4 v\mdot f_1\mdot f_2\mdot v \ell\mdot f_1\mdot v \ell\mdot f_2\mdot v}{2 \ell\mdot p_1 \ell\mdot p_2 \left(p_1\mdot v\right)^2}-\frac{M^4 \left(v\mdot f_1\mdot f_2\mdot v\right)^2}{4 \ell\mdot p_1 \ell\mdot p_2}\Bigg),
	\end{align}  
which is manifestly gauge-invariant as $f^{\mu\nu}_i \equiv p^\mu_i\varepsilon^\nu_i-\varepsilon^\mu_ip^\nu_i$.
For the second graph, the integrand is
\begin{align}
	A_{\rm comp}^{(1,b)}(p_1,p_2,v)&=\int {d^D\ell\over (2\pi)^D } \delta(2Mv\mdot (\ell+p_1)) {1\over \ell^2}\\
	&\times\Big(\sum_{h_{\ell+p_1},h_{\ell-p_2}} A_{3}(p_1+\ell,v) A_{\rm 3}(p_2-\ell,v)A_{\rm GR}(p_1,-\ell-p_1,\ell-p_2,p_2)\Big)\, ,\nn
	\end{align} 
	where $A_{\rm GR}$ is the graviton amplitude. We note that the manifest gauge-invariant form of the integrand can also be obtained by the gauge-invariant double copy form of $A_{\rm GR}$~\cite{Chen:2024gkj}.
	 
 From the glued integrand, we perform the IBP reduction and get
\begin{align}\label{eq:comp1a}
	&A_{\rm comp}^{(1,a)}(p_1,p_2,v)=M^3\left(v_1\mdot f_1\mdot f_2\mdot v_1\right)^2\nn\\
    & \times\left(\frac{\mathcal{K}_3 \left(3-2 y^2\right)}{3 (y-1)^2}+\frac{\mathcal{K}_1 (3-4 y)}{72 (y-1)^2 \omega ^4}+\frac{\mathcal{K}_1 (6-5 y)}{24 (y-1)^2 \omega ^4 \epsilon }-\frac{\mathcal{K}_3 y (31 y-36) \epsilon }{9 (y-1)^2}\right)\nn\\
    &+M^3\left(p_1\mdot f_2\mdot v_1\right)^2 \left(p_2\mdot f_1\mdot v_1\right)^2 \nn\\
    &\times\left(\frac{\mathcal{K}_3 \left(10 y^2+8 y-3\right) \epsilon }{72 (y-1)^4 \omega ^4}-\frac{\mathcal{K}_1 (17 y+4)}{576 (y-1)^4 \omega ^8}+\frac{\mathcal{K}_3 y (y+2)}{12 (y-1)^4 \omega ^4}+\frac{\mathcal{K}_1 (y+2)}{96 (y-1)^4 \omega ^8 \epsilon }\right)\nn\\
   & + M^3 v_1\mdot f_1\mdot f_2\mdot v_1 p_1\mdot f_2\mdot v_1 p_2\mdot f_1\mdot v_1 \nn\\
   &\times\left(\frac{\mathcal{K}_3 \left(13 y^2-29 y+9\right) \epsilon }{9 (y-1)^3 \omega ^2}+\frac{\mathcal{K}_1 (7-2 y)}{72 (y-1)^3 \omega ^6}+\frac{2 \mathcal{K}_3 (y-2) y}{3 (y-1)^3 \omega ^2}+\frac{\mathcal{K}_1 (y-2)}{12 (y-1)^3 \omega ^6 \epsilon }\right)\, ,
\end{align}
and
\begin{align}
	A_{\rm comp}^{(1,b)}(p_1,p_2,v)&=M^3{\mathcal{K}_2\over 2}\Big(-\frac{ \left(y^3-y^2+39 y+9\right) \left(p_1\mdot f_2\mdot v\right)^2 \left(p_2\mdot f_1\mdot v\right)^2}{192 \omega^6 (y-1)^4}\nn\\
	&+\frac{ \left(y^3-23 y^2+27 y+27\right) v\mdot f_1\mdot f_2\mdot v p_1\mdot f_2\mdot v p_2\mdot f_1\mdot v}{48 \omega^4 (y-1)^3}\nn\\
	&-\frac{ \left(y^3-45 y^2+15 y+45\right) \left(v\mdot f_1\mdot f_2\mdot v\right)^2}{48 \omega^2 (y-1)^2}\Big) +(\mathcal{K}_3-\text{term})\, , 
\end{align}
where the $(\mathcal{K}_3-\text{term})$ is the same as the one in eq.\,\eqref{eq:comp1a}. These expressions confirm the result reported in the main text. 

\bibliographystyle{JHEP}
\bibliography{KinematicAlgebra}
\end{document}